\documentclass{article}[a4paper]
\usepackage{array, tabularx, caption, boldline}
\usepackage{graphicx}
\usepackage{makecell}
\usepackage{cellspace}
\usepackage[utf8]{inputenc}
\usepackage{lmodern,textcomp}
\usepackage{hyperref}
\hypersetup{
    colorlinks=true,
    linkcolor=blue,
    filecolor=magenta,      
    urlcolor=blue,
    citecolor=blue
}

\usepackage{float}
\usepackage{cite}
\usepackage{ifthen, mciteplus} 
\usepackage{geometry}
\usepackage{booktabs}
\usepackage{multirow}
\usepackage{amssymb}

\usepackage{url}
\usepackage{placeins}
\newboolean{pdflatex}
\setboolean{pdflatex}{true} 

\newboolean{articletitles}
\setboolean{articletitles}{true} 
\newboolean{uprightparticles}
\setboolean{uprightparticles}{false} 
\newboolean{inbibliography}
\setboolean{inbibliography}{true} 
                             
\usepackage{microtype}
\usepackage{lineno}  
\usepackage{xspace} 
\usepackage{caption} 

\title{Prospects for searches of \bsnunu decays at FCC-ee}
\author{Yasmine Amhis$^{1,2}$, Matthew Kenzie$^{3}$,  Méril Reboud$^{4}$, Aidan R. Wiederhold$^{5\,\dagger}$ }
\date{}

\usepackage{xspace} 
\usepackage{upgreek}
\usepackage{amsmath}
\usepackage{color}

\def\bsnunu{\ensuremath{\bquark\to\squark\neu\neub}\xspace}

\def\BdKSNuNu{\ensuremath{\Bd\to\KS\neu\neub}\xspace}
\def\LbLzNuNu{\ensuremath{\Lb\to\Lz\neu\neub}\xspace}
\def\BdKstNuNu{\ensuremath{\Bd\to\Kstarz\neu\neub}\xspace}
\def\BsPhiNuNu{\ensuremath{\Bs\to\phi\neu\neub}\xspace}
\def\BcDsNuNu{\ensuremath{\Bc\to\Ds\neu\neub}\xspace}
\def\BuKNuNu{\ensuremath{\Bu\to \Kp \neu\neub}\xspace}
\newcommand{\Zbb}{\ensuremath{\Z\to\bquark\bquarkbar}\xspace}
\newcommand{\Zcc}{\ensuremath{\Z\to\cquark\cquarkbar}\xspace}
\newcommand{\Zqq}{\ensuremath{\Z\to\quark\quarkbar}\xspace}
\def\xgboost{\mbox{\textsc{xgboost}}\xspace}

\def\EOS{\texttt{EOS}\xspace}

\def\delphes {\mbox{\textsc{DELPHES}}\xspace}

\def\edmhep{\mbox{\textsc{EDM4hep}}\xspace}

\def\MagUp {\mbox{\em Mag\kern -0.05em Up}\xspace}

\ifthenelse{\boolean{uprightparticles}}%
{
 
 \def\Pgamma      {\ensuremath{\upgamma}\xspace}

 \def\Pnu         {\ensuremath{\upnu}\xspace}                 
                  
 \def\Ppi         {\ensuremath{\uppi}\xspace}

 \def\PDelta      {\ensuremath{\Delta}\xspace}                 
 \def\PXi         {\ensuremath{\Xi}\xspace}                 
 \def\PLambda     {\ensuremath{\Lambda}\xspace}                 
 \def\PSigma      {\ensuremath{\Sigma}\xspace}                 
 \def\POmega      {\ensuremath{\Omega}\xspace}                 
 \def\PUpsilon    {\ensuremath{\Upsilon}\xspace}

 \def\PB      {\ensuremath{\mathrm{B}}\xspace}                 
                  
 \def\PD      {\ensuremath{\mathrm{D}}\xspace}

 \def\PK      {\ensuremath{\mathrm{K}}\xspace}

 \def\PZ      {\ensuremath{\mathrm{Z}}\xspace}                 
                  
 \def\Pb      {\ensuremath{\mathrm{b}}\xspace}                 
 \def\Pc      {\ensuremath{\mathrm{c}}\xspace}                 
 \def\Pd      {\ensuremath{\mathrm{d}}\xspace}                 
 \def\Pe      {\ensuremath{\mathrm{e}}\xspace}

 \def\Pi      {\ensuremath{\mathrm{i}}\xspace}

 \def\Pp      {\ensuremath{\mathrm{p}}\xspace}                 
 \def\Pq      {\ensuremath{\mathrm{q}}\xspace}                 
                  
 \def\Ps      {\ensuremath{\mathrm{s}}\xspace}                 
 \def\Pt      {\ensuremath{\mathrm{t}}\xspace}                 
 \def\Pu      {\ensuremath{\mathrm{u}}\xspace}

 \def\thebaroffset{0.0em}
}
{
 
 \def\Pgamma      {\ensuremath{\gamma}\xspace}

 \def\Pnu         {\ensuremath{\nu}\xspace}                 
                  
 \def\Ppi         {\ensuremath{\pi}\xspace}

 \mathchardef\PDelta="7101
 \mathchardef\PXi="7104
 \mathchardef\PLambda="7103
 \mathchardef\PSigma="7106
 \mathchardef\POmega="710A
 \mathchardef\PUpsilon="7107
                  
 \def\PB      {\ensuremath{B}\xspace}                 
                  
 \def\PD      {\ensuremath{D}\xspace}

 \def\PK      {\ensuremath{K}\xspace}

 \def\PZ      {\ensuremath{Z}\xspace}                 
                  
 \def\Pb      {\ensuremath{b}\xspace}                 
 \def\Pc      {\ensuremath{c}\xspace}                 
 \def\Pd      {\ensuremath{d}\xspace}                 
 \def\Pe      {\ensuremath{e}\xspace}

 \def\Pi      {\ensuremath{i}\xspace}

 \def\Pp      {\ensuremath{p}\xspace}                 
 \def\Pq      {\ensuremath{q}\xspace}                 
                  
 \def\Ps      {\ensuremath{s}\xspace}                 
 \def\Pt      {\ensuremath{t}\xspace}                 
 \def\Pu      {\ensuremath{u}\xspace}

 \def\thebaroffset{0.18em}
}
\newcommand{\offsetoverline}[2][\thebaroffset]{\kern #1\overline{\kern -#1 #2}}%

\makeatletter
\ifcase \@ptsize \relax
  \newcommand{\miniscule}{\@setfontsize\miniscule{4}{5}}
\or
  \newcommand{\miniscule}{\@setfontsize\miniscule{5}{6}}
\or
  \newcommand{\miniscule}{\@setfontsize\miniscule{5}{6}}
\fi
\makeatother

\DeclareRobustCommand{\optbar}[1]{\shortstack{{\miniscule (\rule[.5ex]{1.25em}{.18mm})}
  \\ [-.7ex] $#1$}}

\def\en         {{\ensuremath{\Pe^-}}\xspace}   
\def\ep         {{\ensuremath{\Pe^+}}\xspace}

\def\ellm       {{\ensuremath{\ell^-}}\xspace}
\def\ellp       {{\ensuremath{\ell^+}}\xspace}

\def\neu        {{\ensuremath{\Pnu}}\xspace}
\def\neub       {{\ensuremath{\overline{\Pnu}}}\xspace}

\def\g      {{\ensuremath{\Pgamma}}\xspace}

\def\Z      {{\ensuremath{\PZ}}\xspace}

\def\quark     {{\ensuremath{\Pq}}\xspace}
\def\quarkbar  {{\ensuremath{\overline \quark}}\xspace}
\def\qqbar     {{\ensuremath{\quark\quarkbar}}\xspace}
\def\uquark    {{\ensuremath{\Pu}}\xspace}

\def\dquark    {{\ensuremath{\Pd}}\xspace}

\def\squark    {{\ensuremath{\Ps}}\xspace}

\def\cquark    {{\ensuremath{\Pc}}\xspace}
\def\cquarkbar {{\ensuremath{\overline \cquark}}\xspace}
\def\ccbar     {{\ensuremath{\cquark\cquarkbar}}\xspace}
\def\bquark    {{\ensuremath{\Pb}}\xspace}
\def\bquarkbar {{\ensuremath{\overline \bquark}}\xspace}
\def\bbbar     {{\ensuremath{\bquark\bquarkbar}}\xspace}
\def\tquark    {{\ensuremath{\Pt}}\xspace}
\def\tquarkbar {{\ensuremath{\overline \tquark}}\xspace}

\def\pion   {{\ensuremath{\Ppi}}\xspace}
\def\piz    {{\ensuremath{\pion^0}}\xspace}
\def\pip    {{\ensuremath{\pion^+}}\xspace}
\def\pim    {{\ensuremath{\pion^-}}\xspace}

\def\pimp   {{\ensuremath{\pion^\mp}}\xspace}

\def\hp     {{\ensuremath{h^+}}\xspace}
\def\hm     {{\ensuremath{h^-}}\xspace}

\def\kaon    {{\ensuremath{\PK}}\xspace}

\def\KorKbar {\kern \thebaroffset\optbar{\kern -\thebaroffset \PK}{}\xspace}

\def\Kp      {{\ensuremath{\kaon^+}}\xspace}
\def\Km      {{\ensuremath{\kaon^-}}\xspace}
\def\Kpm     {{\ensuremath{\kaon^\pm}}\xspace}

\def\KS      {{\ensuremath{\kaon^0_{\mathrm{S}}}}\xspace}

\def\Kstarz  {{\ensuremath{\kaon^{*0}}}\xspace}

\def\Kstar   {{\ensuremath{\kaon^*}}\xspace}

\def\D       {{\ensuremath{\PD}}\xspace}

\def\DorDbar {\kern \thebaroffset\optbar{\kern -\thebaroffset \PD}\xspace}
\def\Dz      {{\ensuremath{\D^0}}\xspace}

\def\Dm      {{\ensuremath{\D^-}}\xspace}

\def\Dstar   {{\ensuremath{\D^*}}\xspace}

\def\Dstarz  {{\ensuremath{\D^{*0}}}\xspace}

\def\Dstarm  {{\ensuremath{\D^{*-}}}\xspace}

\def\Ds      {{\ensuremath{\D^+_\squark}}\xspace}
\def\Dsp     {{\ensuremath{\D^+_\squark}}\xspace}

\def\B       {{\ensuremath{\PB}}\xspace}

\def\BorBbar {\kern \thebaroffset\optbar{\kern -\thebaroffset \PB}\xspace}
\def\Bz      {{\ensuremath{\B^0}}\xspace}

\def\Bd      {{\ensuremath{\B^0}}\xspace}

\def\BdorBdbar {\kern \thebaroffset\optbar{\kern -\thebaroffset \Bd}\xspace}
\def\Bu      {{\ensuremath{\B^+}}\xspace}

\def\Bp      {{\ensuremath{\Bu}}\xspace}

\def\Bs      {{\ensuremath{\B^0_\squark}}\xspace}

\def\BsorBsbar {\kern \thebaroffset\optbar{\kern -\thebaroffset \Bs}\xspace}
\def\Bc      {{\ensuremath{\B_\cquark^+}}\xspace}

\def\Y#1S{\ensuremath{\PUpsilon{(#1S)}}\xspace}

\def\proton      {{\ensuremath{\Pp}}\xspace}

\def\porpbar     {\kern \thebaroffset\optbar{\kern -\thebaroffset \proton}\xspace}
\def\Lz          {{\ensuremath{\PLambda}}\xspace}

\def\LorLbar     {\kern \thebaroffset\optbar{\kern -\thebaroffset \PLambda}\xspace}

\def\Lb           {{\ensuremath{\Lz^0_\bquark}}\xspace}

\def\BF         {{\ensuremath{\mathcal{B}}}\xspace}

\def\to                 {\ensuremath{\rightarrow}\xspace}

\def\CP                {{\ensuremath{C\!P}}\xspace}

\def\AT#1     {\ensuremath{A_{\mathrm{T}}^{#1}}\xspace}           

\def\C#1      {\ensuremath{\mathcal{C}_{#1}}\xspace}                       
\def\Cp#1     {\ensuremath{\mathcal{C}_{#1}^{'}}\xspace}                    
\def\Ceff#1   {\ensuremath{\mathcal{C}_{#1}^{\mathrm{(eff)}}}\xspace}        
\def\Cpeff#1  {\ensuremath{\mathcal{C}_{#1}^{'\mathrm{(eff)}}}\xspace}       
\def\Ope#1    {\ensuremath{\mathcal{O}_{#1}}\xspace}                       
\def\Opep#1   {\ensuremath{\mathcal{O}_{#1}^{'}}\xspace}                    

\newcommand{\bra}[1]{\ensuremath{\langle #1|}}             
\newcommand{\ket}[1]{\ensuremath{|#1\rangle}}              

\newcommand{\nospaceunit}[1]{\ensuremath{\text{#1}}}       
\newcommand{\aunit}[1]{\ensuremath{\text{\,#1}}}       

\newcommand{\tev}{\aunit{Te\kern -0.1em V}\xspace}
\newcommand{\gev}{\aunit{Ge\kern -0.1em V}\xspace}
\newcommand{\mev}{\aunit{Me\kern -0.1em V}\xspace}
\newcommand{\kev}{\aunit{ke\kern -0.1em V}\xspace}
\newcommand{\ev}{\aunit{e\kern -0.1em V}\xspace}
\newcommand{\mevc}{\ensuremath{\aunit{Me\kern -0.1em V\!/}c}\xspace}
\newcommand{\gevc}{\ensuremath{\aunit{Ge\kern -0.1em V\!/}c}\xspace}
\newcommand{\mevcc}{\ensuremath{\aunit{Me\kern -0.1em V\!/}c^2}\xspace}
\newcommand{\gevcc}{\ensuremath{\aunit{Ge\kern -0.1em V\!/}c^2}\xspace}

\def\mm   {\aunit{mm}\xspace}

\def\mum  {\ensuremath{\,\upmu\nospaceunit{m}}\xspace}

\newcommand{\chisq}{\ensuremath{\chi^2}\xspace}

\def\gsim{{~\raise.15em\hbox{$>$}\kern-.85em
          \lower.35em\hbox{$\sim$}~}\xspace}
\def\lsim{{~\raise.15em\hbox{$<$}\kern-.85em
          \lower.35em\hbox{$\sim$}~}\xspace}

\def\evtgen     {\mbox{\textsc{EvtGen}}\xspace}

\def\photos     {\mbox{\textsc{Photos}}\xspace}

\def\pythia     {\mbox{\textsc{Pythia}}\xspace}

\def\tell1  {TELL1\xspace}
\def\ukl1   {UKL1\xspace}

\newcommand{\ie}{\mbox{\itshape i.e.}\xspace}

\newcommand{\etc}{\mbox{\itshape etc.}\xspace}

\newcommand{\vs}{\mbox{\itshape vs.}\xspace}

\def\BF {{\ensuremath{\mathcal{B}}}\xspace}

\begin{document}

\begin{flushright}
    \href{https://doi.org/10.1007/JHEP01(2024)144}{DOI:10.1007/JHEP01(2024)144}\\
    \href{https://doi.org/10.17181/6k4q7-veh06}{DOI: 10.17181/6k4q7-veh06} \\
    EOS-2023-04\\
    IPPP/23/51 
\end{flushright}
\vspace*{3\baselineskip}

{\let\newpage\relax\maketitle}

\begin{center}{\footnotesize \it
\noindent
$^{1}$Universit\'e Paris-Saclay, CNRS/IN2P3, IJCLab, Orsay, France \\
$^{2}$European Organization for Nuclear Research (CERN), Geneva, Switzerland \\
$^{3}$Cavendish Laboratory, University of Cambridge, Cambridge, UK \\
$^{4}$IPPP, Durham University, Durham, UK \\
$^{5}$Department of Physics, University of Warwick, Coventry, UK \\
$^{\dagger}$Corresponding Author
 }

\vspace{0.5cm}

{\footnotesize
{{Email:~}}{\bf\color{blue} yasmine.sara.amhis@cern.ch, matthew.kenzie@cern.ch, \\ 
merilreboud@gmail.com,  aidan.richard.wiederhold@cern.ch}
}
\end{center}
\smallskip

\begin{center}
Published in JHEP \textbf{01}~(2024)~144
\end{center}
\smallskip

\hrule
\begin{abstract}\noindent
We investigate the physics reach and potential for the study of various decays involving a \bsnunu transition at the Future Circular Collider running electron-positron collisions at the $Z$-pole (FCC-ee). 
Signal and background candidates, which involve inclusive $Z$ contributions from $b\bar{b}$, $c\bar{c}$ and $uds$ final states, are simulated for a proposed multi-purpose detector. Signal candidates are selected using two Boosted Decision Tree algorithms. 
We determine expected relative sensitivities of $0.53\%$, $1.20\%$, $3.37\%$ and $9.86\%$ for the branching fractions of the \BdKstNuNu, \BsPhiNuNu, \BdKSNuNu and \LbLzNuNu decays, respectively.  
In addition, we investigate the impact of detector design choices related to particle-identification and vertex resolution.
The phenomenological impact of such measurements on the extraction of Standard Model and new physics parameters is also studied.

\end{abstract}

\vspace*{0.5em}

\hrule

\clearpage 

\tableofcontents
\clearpage


\section{Introduction}
\label{sec:introduction}

Flavor Changing Neutral Current (FCNC) processes are sensitive probes of New Physics (NP) effects since they are both loop- and CKM-suppressed in the Standard Model (SM).
Over the past several years, an enormous effort has been made at the LHC~\cite{LHCb:2015svh,LHCb:2016ykl,LHCb:2021zwz,LHCb:2014cxe,LHCb:2020lmf,LHCb:2014auh,LHCb:2016due,CMS:2017rzx,CMS:2018qih,ATLAS:2018gqc,LHCb:2022qnv} and the $B$-factories~\cite{BaBar:2012mrf,Belle:2016fev,BELLE:2019xld,Belle:2019oag} to precisely measure decays involving a $b\to s\ell\ell$ transition.
However, a challenge which prohibits full exploitation of this data is precise knowledge of the SM predictions of the relevant observables, which are in most cases plagued by hadronic uncertainties, see e.g.~Ref~\cite{Ciuchini:2022wbq,Gubernari:2022hxn}. \\

The main interest in studying the decays involving a \bsnunu transition is that they are theoretically cleaner than their counterparts with charged leptons~\cite{Altmannshofer:2009ma,Buras:2014fpa,Becirevic:2023aov}.
Charm loops do not contribute to \bsnunu decays, which are, baring weak annihilation effects that we will discuss, dominated by short-distance effects that have been precisely computed, including subleading QCD and electroweak corrections~\cite{Kamenik:2009kc,Buchalla:1993bv,Buchalla:1998ba,Misiak:1999yg,Brod:2010hi}.
The only remaining theoretical uncertainties originate from knowledge of the CKM factor $V^{}_{tb}V_{ts}^\ast$, which can be determined using CKM unitarity~\cite{Buras:2014fpa}, as well as the relevant local form-factors, which can be computed by means of numerical simulations of QCD on the lattice~\cite{FlavourLatticeAveragingGroupFLAG:2021npn}.
Recently, it has also been shown that one could probe \CP-violating effects via time-dependent analysis of \BdKSNuNu and \BdKstNuNu decays~\cite{Descotes-Genon:2022gcp}.

Another motivation to study the \bsnunu transition is its sensitivity to NP contributions.
Most importantly, \bsnunu observables allow us to probe effective operators with couplings to $\nu_\tau$, which are related by $SU(2)_L$ gauge invariance to operators with left-handed $\tau$-leptons~\cite{Buchmuller:1985jz,Bause:2020auq,Bause:2021cna}.
These operators are poorly constrained at low-energies due to the experimental difficulty of probing decays involving a $b\to s\tau\tau$ transition~\cite{LHCb:2017myy}.
Furthermore, \bsnunu observables can be related by gauge invariance to the hints of lepton-flavor-universality violation in the $b\to c\tau \nu$ transition, which are still to be clarified, cf.~e.g.~\cite{Angelescu:2021lln,Cornella:2021sby}. \\
 
Experimentally, the first evidence for the $\Bp\to \Kp \nu \bar{\nu}$ decay has been found recently by the Belle II collaboration~\cite{Belle-II:2023esi} with a significance of $3.5\sigma$.
Interestingly, the measured branching ratio $\mathcal{B}(\Bp\to \Kp \nu \bar{\nu}) =  (2.3 \pm 0.7) \times 10^{-5}$ exceeds the Standard Model prediction by $2.7\sigma$ \cite{Belle-II:2023esi}.
The Belle II experiment is also working on $\Bd\to \Kstarz \neu \neub$ decays, for which only upper limits have been obtained so far.
In the future, they are expected to measure the corresponding branching fractions with $\mathcal{O}(10\%)$ experimental precision with $50~\mathrm{ab}^{-1}$ of data~\cite{Belle-II:2018jsg}.
These measurements are particularly challenging experimentally due to the missing energy of the neutrinos, and are consequently ideally suited to the clean environment of an electron-positron collider.
$Z$-boson factories such as the Future Circular Collider running at the $Z$-pole (FCC-ee) offer a unique opportunity to study these decays in the future and to significantly improve on the precision that will be achieved by Belle II. \\

In this paper, we perform a sensitivity study of various \bsnunu decays at FCC-ee.
These include the \BdKSNuNu and \BdKstNuNu modes, which are accessible at Belle II running at the $\Upsilon(4S)$ resonance, but also \BsPhiNuNu and \LbLzNuNu which can only be measured in a Tera-Z experiment such as FCC-ee.
In this work we do not study \BuKNuNu or \BcDsNuNu decays however measurements of these at the FCC-ee would certainly be possible and can be a topic of further feasibility and detector requirement studies.
The \BuKNuNu decay mode is considerably more challenging than those we present here because the companion \kaon is stable and therefore has no decay vertex. 
This mode would require specialised reconstruction that was considered beyond the scope for this study. 
The \BcDsNuNu decay also requires a slightly more complicated analysis due to the wide variety of \Dsp final states, although the three-prong vertex would perhaps give a boost in the relative precision.
Furthermore the \Bc hadronisation fraction is $\mathcal{O}(100)$ times smaller than that of \Bd mesons so will inherently be measured with much smaller precision.
For the decays that are studied herein we employ a similar strategy to Ref.~\cite{Amhis:2021cfy}, in which we exploit the relatively large imbalance of missing energy between the signal hemisphere (which contains two neutrinos) and the non-signal hemisphere. 
We then train a sequence of two boosted decision trees (BDTs) to distinguish between signal-like and background-like events, the first focusing on global event information and the second on specific candidate information.
We use these two BDTs to optimise selection cuts and thus estimate the expected sensitivity to the relevant signal. \\

The remainder of this paper is organized as follows.
Section~\ref{sec:theory} describes SM predictions of branching fractions and form factors in \bsnunu transitions.
Section~\ref{sec:experimental-setup} describes the experimental environment of the FCC and the IDEA detector.
Section~\ref{sec:analysis} describes the analysis performed and provides results for the sensitivity estimates, along with some discussion on detector design implications.
Section~\ref{sec:interpretation} provides the interpretation of the sensitivity estimates in terms of SM parameters and of the relevant effective field theory Wilson coefficients.

\section{SM predictions}
\label{sec:theory}

The Weak Effective Theory (WET) Hamiltonian describing the \bsnunu transition can be written as
\begin{equation}
    \label{eq:th:Heff}
    \mathcal{H}_\mathrm{eff}^{sb\nu\nu}
        = -\frac{4 G_F}{\sqrt{2}} \lambda_t \sum_i \mathcal{C}_i \mathcal{O}_i + \text{h.c.},
\end{equation}
where $G_F$ denotes the Fermi constant and $\lambda_t = V_{tb}^{\phantom{*}} V_{ts}^*$ is the CKM factor.
In the SM, the only non-zero Wilson coefficient, $\mathcal{C}_L$, is associated to the operator
\begin{equation}
    \mathcal{O}_L^{\nu_i,\nu_j} =
        \frac{e^2}{16 \pi^2} \big( \bar{s}_L \gamma_\mu b_L \big) \big( \bar{\nu}_i \gamma^\mu (1-\gamma_5) \nu_j \big),
\end{equation}
where $\left. \mathcal{C}_L^{\nu_i,\nu_j} \right|_\mathrm{SM} = \delta_{ij} C_L^\mathrm{SM}$, with
\begin{equation}
C_L^\mathrm{SM} = -\frac{1.462(17)(2)}{\sin^2 \theta_W},
\end{equation}
where NLO QCD corrections and NNLO electroweak contributions are taken into account~\cite{Buchalla:1998ba, Misiak:1999yg, Brod:2010hi}.
Using $\sin^2 \theta_W = 0.23141(4)$~\cite{PDG2022}, one gets $C_L^\mathrm{SM} = -6.32(7)$, with the dominant source of uncertainty due to higher-order QCD corrections.
These uncertainties are negligible when compared to the theory uncertainties that will be discussed below.\\

Several decay modes of $b$-hadrons can be induced by the effective Hamiltonian in Eq.~\eqref{eq:th:Heff}.
The only ones accessible at Belle II are $B\to K \nu \bar{\nu}$ and $B\to K^\ast \nu \bar{\nu}$, with mesons that can be either electrically charged or electrically neutral~\cite{Belle-II:2018jsg}.
All the other modes cannot be measured in any of the running and future experiments, except for FCC-ee, which, as we will show, can additionally access $B_s\to \phi \nu \bar{\nu}$ and $\Lb \to \Lz^{(*)} \nu \bar{\nu}$.
In what follows, we will limit ourselves to the decays involving neutral mesons, namely \BdKSNuNu, \BdKstNuNu, \BsPhiNuNu and \LbLzNuNu, collectively referred to as $B\to Y\neu\neub$ throughout this paper, for two reasons.
First, they are not affected by weak annihilation contributions~\cite{Kamenik:2009kc} which makes them theoretically cleaner.
Second, they are experimentally easier to probe as the decay vertex of the neutral hadron into charged tracks is reconstructible.

The relevant decay rates can be written in the SM as follows \cite{Buras:2014fpa,Chen:2000mr},
\begin{align}
    \label{eq:th:BRBdKS}
    \frac{d\BF(\BdKSNuNu)_\mathrm{SM}}{dq^2} & =
        3 \, \tau_{\Bd} |N_{\Bd}|^2 |C_L^\mathrm{SM}|^2 |\lambda_t|^2 \rho^\KS_+\,, \\[0.35em]
    \frac{d\BF(\BdKstNuNu)_\mathrm{SM}}{dq^2} & =
        3 \, \tau_{\Bd} |N_{\Bd}|^2 |C_L^\mathrm{SM}|^2 |\lambda_t|^2 (\rho^\Kstarz_{A_1} + \rho^\Kstarz_{A_{12}} + \rho^\Kstarz_V)\,, \\[0.35em]
    \frac{d\BF(\BsPhiNuNu)_\mathrm{SM}}{dq^2} & =
        3 \, \tau_{\Bs} |N_{\Bs}|^2|C_L^\mathrm{SM}|^2 |\lambda_t|^2 (\rho^\phi_{A_1} + \rho^\phi_{A_{12}} + \rho^\phi_V)\,, \\[0.35em]
    \label{eq:th:BRLbLz}
    \frac{d\BF(\LbLzNuNu)_\mathrm{SM}}{dq^2} & =
        3 \, \tau_{\Lb} |N_{\Lb}|^2 |C_L^\mathrm{SM}|^2 |\lambda_t|^2 (\rho^\Lz_{f_\perp^V} + \rho^\Lz_{f_\perp^A} + \rho^\Lz_{f_0^V} + \rho^\Lz_{f_0^A})\,,
\end{align}
where
\begin{equation}
    N_{B_q} = \frac{G_F \alpha_\mathrm{em}}{16 \pi^2} \sqrt{\frac{m_{B_q}}{3 \pi}}.
\end{equation}
In the above equations, $\rho_i\equiv\rho_i(q^2)$ are functions of the hadronic form factors defined in Appendix~\ref{app:sec:from-factors}, 
\begin{align}
    \rho^\KS_+ & = \frac{\lambda^{3/2}}{2 m_\Bd^4}
        \left( f^K_+(q^2) \right)^2, \\[0.35em]
    \rho^\Kstarz_V  & = \frac{2 \, q^2 \lambda^{3/2}}{(m_\Bd + m_\Kstarz) m_\Bd^4}
        \left( V^\Kstar(q^2) \right)^2, \\[0.35em]
    \rho^\Kstarz_{A_1}  & = \frac{2 \, q^2 \lambda^{1/2} (m_\Bd + m_\Kstarz)^2}{m_\Bd^4}
        \left( A_1^\Kstar(q^2) \right)^2, \\[0.35em]
    \rho^\Kstarz_{A_{12}}  & = \frac{64 \, m_\Kstarz^2 \lambda^{1/2}}{m_\Bd^2}
        \left( A_{12}^\Kstar(q^2) \right)^2, \\[0.35em]
    \rho^\Lz_{f_\perp^{V/A}}  & = \frac{32 \, q^2 \lambda^{1/2} ((m_\Lb \mp m_\Lz)^2 - q^2)}{m_\Lb^4}
        \left( f_\perp^{V/A}(q^2) \right)^2, \\[0.35em]
    \rho^\Lz_{f_0^{V/A}}  & = \frac{16 \, \lambda^{1/2} (m_\Lb \pm m_\Lz)^2 ((m_\Lb \mp m_\Lz)^2 - q^2)}{m_\Lb^4}
        \left( f_0^{V/A}(q^2) \right)^2,
\end{align}
where $\lambda \equiv \lambda(q^2, m_1^2, m_2^2) = (q^2-(m_1-m_2)^2) (q^2-(m_1+m_2)^2)$.
Moreover, the expressions for $\Bs\to\phi$ are obtained from $\Bd\to\Kstarz$ via trivial replacements.
An angular analysis of these decays offers access to one additional observable for $\BdKstNuNu$ and $\BsPhiNuNu$ and two additional observables for $\LbLzNuNu$.
Following Refs.~\cite{Altmannshofer:2009ma, Buras:2014fpa, Felkl:2021uxi, Das:2017ebx} we define the mesonic longitudinal polarisation fractions as
\begin{align}
    F_L(\BdKstNuNu)_\mathrm{SM} &= \frac{\rho^\Kstarz_{A_{12}}}{\rho^\Kstarz_{A_1} + \rho^\Kstarz_{A_{12}} + \rho^\Kstarz_V}\,,\\[0.35em]
    F_L(\BsPhiNuNu)_\mathrm{SM} &= \frac{\rho^\phi_{A_{12}}}{\rho^\phi_{A_1} + \rho^\phi_{A_{12}} + \rho^\phi_V}\,.
\end{align}
The $\LbLzNuNu$ longitudinal polarisation fractions and hadronic forward backward asymmetry are derived from Ref.~\cite{Boer:2014kda} and read
\begin{align}
        F_L(\LbLzNuNu)_\mathrm{SM}  &= \frac{\rho^\Lz_{f_0^V} + \rho^\Lz_{f_0^A}}{\rho^\Lz_{f_\perp^V} + \rho^\Lz_{f_\perp^A} + \rho^\Lz_{f_0^V} + \rho^\Lz_{f_0^A}}\,,\\[0.35em]
        A_\mathrm{FB}^\Lz(\LbLzNuNu)_\mathrm{SM} &= \frac{\alpha}{2} \frac{\tilde{\rho}^\Lz_\perp + \tilde{\rho}^\Lz_0}{\rho^\Lz_{f_\perp^V} + \rho^\Lz_{f_\perp^A} + \rho^\Lz_{f_0^V} + \rho^\Lz_{f_0^A}}\,,
\end{align}
where $\alpha$ is the parity-violating decay parameter defined in Ref.~\cite{Boer:2014kda} and we used
\begin{align}
    \tilde{\rho}^\Lz_\perp  & = \frac{32 \, q^2 \lambda^{1/2} ((m_\Lb \mp m_\Lz)^2 - q^2)}{m_\Lb^4} \, f_\perp^V(q^2) f_\perp^A(q^2), \\[0.35em]
    \tilde{\rho}^\Lz_0  & = \frac{16 \, \lambda^{1/2} (m_\Lb \pm m_\Lz)^2 ((m_\Lb \mp m_\Lz)^2 - q^2)}{m_\Lb^4} \, f_0^V(q^2) f_0^A(q^2).
\end{align}

\noindent There are two main sources of uncertainties in the prediction of these decays rates: (i) the value of the CKM product $\lambda_t$ and (ii) the hadronic form factors that need to be determined non-perturbatively, which will be discussed in the following. 

The usual strategy to determine $\lambda_t$ is to use the unitarity of the CKM matrix to relate it to $|V_{cb}|$~\cite{Buras:2014fpa}.
However, the current discrepancy between the inclusive and exclusive determinations of $|V_{cb}|$ introduces an ambiguity in the values that could be taken, see e.g.~Ref.~\cite{Becirevic:2023aov} for a recent discussion.
An alternative is to extract $|\lambda_t|$ from the mass-difference in the $B_s-\overline{B_s}$ system using the product $f_{B_s} \sqrt{\hat{B}_s}$ of the $B_s$ decay constant and bag parameter computed on the lattice~\cite{Buras:2021nns,Buras:2022wpw}.
However, there is currently a disagreement between the determinations with $2+1$ and $2+1+1$ dynamical flavors~\cite{FlavourLatticeAveragingGroupFLAG:2021npn}, which leads again to an ambiguity.
For the sake of definiteness, we will consider the value $|\lambda_t|= (39.3 \pm 1.0)\times 10^{-3}$ based on $|V_{cb}|= (40.0 \pm 1.0)\times 10^{-3}$ extracted from $B\to D \ell \nu$ decays, which has a relative uncertainty of $\approx 2.5\%$~\cite{FlavourLatticeAveragingGroupFLAG:2021npn}.
However, it is clear that this puzzle needs to be solved by a combined theoretical and experimental effort to match the experimental precision foreseen at FCC-ee.
In the phenomenological analysis of Sec.~\ref{sec:interpretation}, we will also consider a hypothetical uncertainty of $\approx 1.5\%$ which is quoted for exclusive $|V_{cb}|$ determinations at Belle II with $50~\mathrm{ab}^{-1}$~\cite{Belle-II:2018jsg}.

Regarding the hadronic form factors, the most reliable determinations are those based on numerical simulations of QCD on the lattice (LQCD).
However, these results are only available for a few decay channels and only for large $q^2$-values.
The SM predictions of the branching fractions thus rely on extrapolations of the form factors to the entire physical region, which are based on specific parameterisations.
An alternative method, discussed in Sec.~\ref{sec:interpretation}, consists of extracting ratios of form factors directly from the data, to guide the extrapolation at low $q^2$.

In our phenomenological analysis, performed using the open-source \EOS software~\cite{EOSAuthors:2021xpv} version v1.0.10~\cite{EOS:v1.0.10}, we will consider two sets of form factors:
\begin{description}
    \item[2023] For the mesonic modes $B\to K^{(*)}$ and $B_s\to\phi$, we follow the approach of Ref.~\cite{Gubernari:2023puw} and parametrise the form factors with simplified series expansions~\cite{Bharucha:2010im}.
    We use the LQCD $B\to K$ inputs of the FNAL\-/\-MILC~\cite{Bailey:2015dka} and HPQCD~\cite{Parrott:2022rgu} collaborations.
    The $B \to K^\ast$ and $B_s \to \phi$ transitions are fitted on the LQCD inputs of Ref.~\cite{Horgan:2013pva} and the Light-Cone Sum Rules (LCSR) estimations of Refs.~\cite{Gubernari:2018wyi,Gubernari:2020eft}.
    For the baryonic mode we follow Ref.~\cite{Blake:2022vfl} which uses the LQCD inputs of Ref.~\cite{Detmold:2016pkz}.\\
    The predictions based on these inputs are quoted in Table~\ref{tab:SMpred}.
    The uncertainties due to the form factors amounts to $5\%$ for the $B\to K$ transition and $10\%$ for the other transitions.

    \item[Future] The predictions will need to be considerably improved to match the experimental precision foreseen at FCC-ee.
    This would be particularly challenging for transitions featuring resonances, as they are notably harder to predict, especially if these resonances are broad.
    For the purposes of this analysis, we assume that the uncertainties will be reduced by a factor of ten over the coming decades.
    This scenario only serves as a reference to make the phenomenological analysis realistic.
    The uncertainties due to the form factors would therefore amount to less than a percent for $B\to K$ and $\sim 1\%$ for the other transitions.
\end{description}

\begin{table}[h!]
\renewcommand{\arraystretch}{1.7}
	\centering
	\begin{tabular}{@{}l c c c@{}}
        \toprule
        Decay mode & $\mathcal{B}/|\lambda_t|^2 \, [10^{-3}]$ & $\mathcal{B}\, [10^{-6}]$ & Ref.  \\
        \midrule
        \BdKSNuNu  & $1.33 \pm 0.04$ & $2.02 \pm 0.12$ & \cite{Gubernari:2023puw, Becirevic:2023aov}\\
        \BdKstNuNu & $5.13 \pm 0.51$ & $7.93 \pm 0.89$ & \cite{Gubernari:2023puw}\\
        \BsPhiNuNu & $6.31 \pm 0.67$ & $9.74 \pm 1.15$ & \cite{Gubernari:2023puw}\\
        \LbLzNuNu  & $5.55 \pm 0.56$ & $8.57 \pm 0.97$ & \cite{Blake:2022vfl}\\
        \bottomrule
	\end{tabular}
    \caption{\label{tab:SMpred}
        Current SM prediction for integrated decay rates summed over the three neutrino flavors.
        In the second row we used the exclusive determination of $|V_{cb}|= (40.0 \pm 1.0)\times 10^{-3}$, which yields $|\lambda_t|= (39.3 \pm 1.0)\times 10^{-3}$, as described in the text.
    }
\end{table}

\section{Experimental environment}
\label{sec:experimental-setup}

For our experimental analysis, we follow much of the procedure developed and outlined in Ref.~\cite{Amhis:2021cfy}.
Here we give a brief description of the collider and detector environment that has been assumed for this study.

\subsection{FCC-ee}

The proposed Future Circular Collider (FCC)~\cite{FCC:2018byv} is the next generation state-of-the-art particle research facility.
The ongoing FCC feasibility study is investigating the benefits and physics reach of such a machine which would be built in a new 80~--~100~km tunnel, near CERN, with capabilities of running in successive stages of $e^+e^-$, $e$-$p$ or $p$-$p$ mode.
The \ep\en machine (FCC-ee)~\cite{FCC:2018evy} would run at centre-of-mass-energies, $\sqrt{s}$, in the range between 91\gev (\ie the Z-pole) and 365\gev (\ie the \tquark\tquarkbar threshold).
FCC-ee offers unpredecented opportunity to study every known particle of the SM in exquisite detail. 
Beyond its capabilities as an electroweak precision machine there is scope for world's-best measurements in the beauty (\bquark-quark), charm (\cquark-quark) and tau ($\tau$-lepton) sectors with the vast statistics anticipated to be taken at the Z-pole. 
This so called ``Tera-$Z$" run would produce $\mathcal{O}(10^{12})$ $Z$-bosons per experiment, which have a high branching fraction to both \bquark\bquarkbar (0.15) and \cquark\cquarkbar (0.12) pairs~\cite{PDG2022}.
In contrast with other proposed future colliders, such as the ILC, the low-energy operation of the FCC-ee allows for greater instantaneous luminosity by a factor of $\mathcal{O}(100)$.
The result is FCC-ee data samples orders of magnitude larger than could be acquired at the ILC and consequently allows for considerably more precise measurements \cite{Bambade:2019fyw}.
Another advantage of a circular, as opposed to linear,  collider layout is that collisions can be delivered to multiple interaction regions simultaneously, which allows for a variety of different detector design choices.

\subsection{Detector Response}

Monte-Carlo (MC) event samples are used to simulate the response of the detector to various different physics processes. 
The procedure for event generation and simulation of the detector response is identical to that described in Ref.~\cite{Amhis:2021cfy}.
In summary, events are generated under nominal FCC-ee conditions using \pythia~\cite{Sjostrand:2014zea}, with unstable particles decayed using \evtgen~\cite{Lange:2001uf} and final-state radiation generated by \photos~\cite{Davidson:2010ew}.
The detector configuration under consideration is the Innovative Detector for Electron-positron Accelerators (IDEA) concept.
It consists of a silicon pixel vertex detector, a large-volume extremely-light short-drift wire chamber surrounded by a layer of silicon micro-strip detectors, a thin low-mass superconducting solenoid coil, a pre-shower, a dual-readout calorimeter, and muon chambers within the magnet return yoke~\cite{FCC:2018evy}.
The detector response is simulated using the \delphes package with the configuration card in Ref.~\cite{helsens_clement_2021_4817870} interfaced to the common \edmhep data format~\cite{valentin_volkl_2021_4785063}.


\subsection{Simulation Samples}

Our study exploits various different MC simulation samples used to mimic the expected signal and background distributions at FCC-ee. 
We make use of inclusive samples of $\Z\to\bbbar$, $\Z\to\ccbar$ and $\Z\to\qqbar$ (where $q$ is one of the light quarks, $q\in\{\uquark, \dquark, \squark\}$) as proxies for the total expected background. 
We then make use of dedicated exclusive samples for each of the signal modes under study, namely the \BdKstNuNu, \BsPhiNuNu, \BdKSNuNu and \LbLzNuNu decays.
The simulated samples contain an admixture of both \bquark-hadron flavours \ie charge-conjugation is implied throughout. 
The \Kstarz resonance is assumed to be pure vector $K^{*}(892)^0 \to \Kp\pim$ and the $\phi$ resonance is assumed to be pure vector $\phi(1020) \to \Kp \Km$.

\begin{figure}
    \centering
    \includegraphics[width=0.49\textwidth]{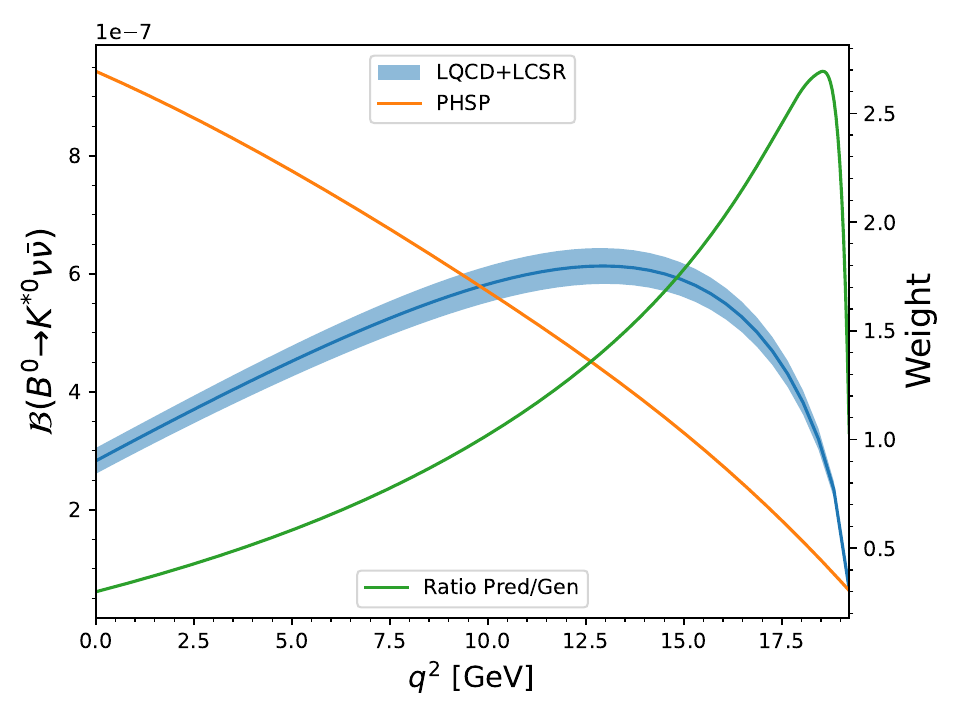}
    \includegraphics[width=0.49\textwidth]{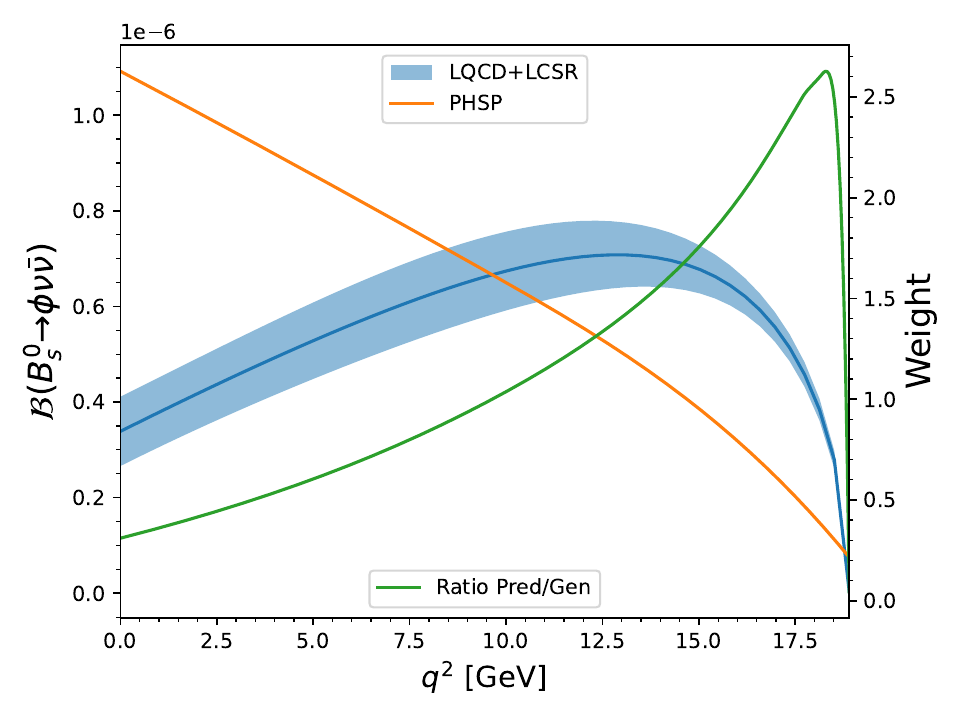}
    \caption{A comparison between the generated $q^2$ distribution (orange line) and the theory prediction provided in this paper (blue line and band) along with their ratio (green line) which is used to reweight the simulation samples in our analysis, for the \BdKstNuNu decay (left) and \BsPhiNuNu decay (right).}
    \label{fig:q2_reweight}
\end{figure}

The signal decays are simulated using the \texttt{PHSP} \texttt{EvtGen} model which generates the $B$ candidate decay children uniformly distributed in phase space.
This does not accurately simulate the correct momentum transfer distribution in these decays.
Consequently, we reweight our simulation samples using the MC truth invariant mass of the neutrino pair, $q^2$, and the model predictions provided in Sec.~\ref{sec:theory}.
A comparison between the \texttt{PHSP} and theory prediction (LQCD+LCSR) for the $q^2$ distribution is shown in Fig.~\ref{fig:q2_reweight} along with their ratio which is used for the reweighting of the simulation samples.

\subsection{Analysis framework and implementation}

We make use of the same basic analysis framework as deployed in Ref.~\cite{Amhis:2021cfy}.
Our nominal analysis strategy (variations on these assumptions are further discussed below) assumes:
\begin{itemize}
    \item \textbf{Perfect vertex seeding.} Whilst we take into account that vertex positions are not perfectly known, via the tracking system resolution, we assume that vertices can be perfectly seeded. In other words we always match the reconstructed vertex to the simulated vertex. The impact of this assumption is studied further below. High precision vertex finding will be a crucial aspect of the detector design to maximise the physics reach for \bsnunu.
    \item \textbf{Perfect particle identification.} We assume that the detector will have perfect discrimination between kaons and pions (and indeed protons and other species). This is particularly relevant for broader resonances that have both kaon and pions in the final state (for example the \Kstarz). The impact of this assumption is studied in further detail below in which we investigate the sensitivity at different values of the kaon-pion separation power.
\end{itemize}

Furthermore, due to the additional complexity required in reconstructing neutral final states, such as $\KS\to\pip\pim$ and $\Lz\to p \pim$, which fly some distance in the detector before producing charged tracks, we do not yet fully reconstruct these modes. 
We instead chose to focus on the modes which decay promptly, \ie with $\Kstarz\to\Kp\pim$ and $\phi\to\Kp\Km$, and make sensitivity projections for the modes with neutrals based on assumptions about the neutral reconstruction. 
Reconstruction of neutral \KS and \Lz candidates has recently been developed for the IDEA detector at FCC-ee but was not available in time for our studies.
A full study which includes neutral reconstruction will come at a later date.

\section{Analysis}
\label{sec:analysis}

In order to obtain an estimate for the expected sensitivity to the various \bsnunu decays under consideration, we optimise a two-stage selection procedure based on Boosted Decision Trees (BDTs). 
These are trained to distinguish between the signal candidates of interest and the inclusive backgrounds from \Zbb, \Zcc and \Zqq, for $q\in\{\uquark,\dquark,\squark\}$.

One of the key signatures of the signal decays is the presence of large missing energy in the direction of the $B$ meson candidate due to the two neutrinos in the final state. 
Consequently a typical signal event will have a relatively large imbalance of missing energy between the signal side of the \Zbb event and the non-signal side. 
For a typical \Zbb background event any missing energy will be approximately the same on both sides. 
In order to determine the imbalance between the signal-side and the non-signal-side we divide events (on a per-event basis) into two hemispheres, each respectively corresponding to one of the two \bquark-quarks produced from the \Z decay.

\begin{figure}
    \centering \includegraphics[width=0.6\textwidth]{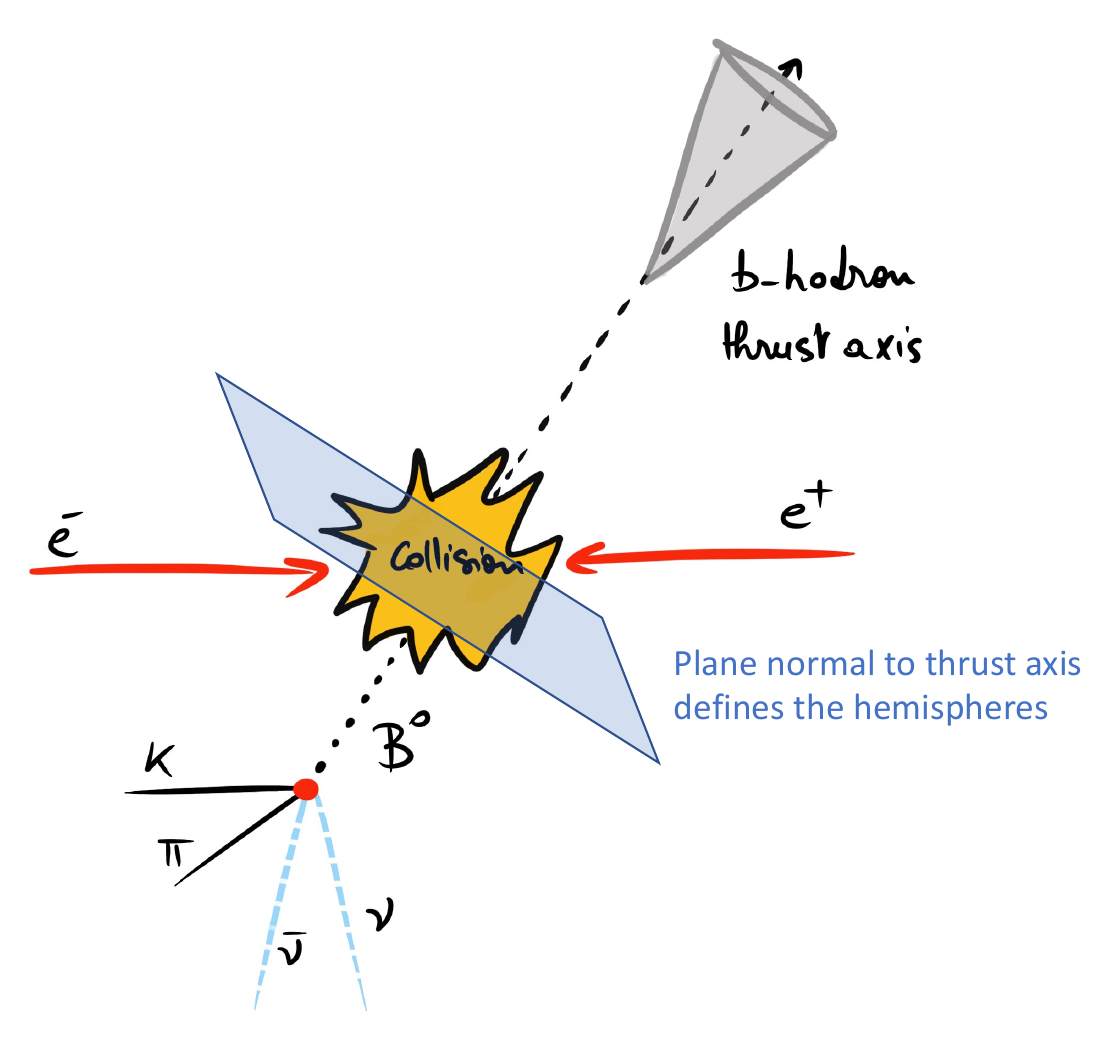}
    \caption{A pictorial representation of the definition of the thrust axis and the two event hemispheres for a \BdKstNuNu event.}
    \label{fig:hemispheres}
\end{figure}

The hemispheres, pictorially represented in Fig.~\ref{fig:hemispheres}, are defined using the plane normal to the thrust axis, which is defined by the unit vector, $\hat{\mathbf{n}}$, that minimises,
\begin{equation}
    T = \frac{\sum_i | \mathbf{p}_i \cdot \hat{\mathbf{n}} |}{\sum_i |\mathbf{p}_i| },
\end{equation}
where $\mathbf{p}_i$ is the momentum vector of the $i^\text{th}$ reconstructed particle in the event.
This thrust axis provides a measure of the direction of the quark pair produced from the \Z decay.
Reconstructed particles from each event are then assigned to either hemisphere depending on the angle, $\theta$, between their momentum vector and the thrust axis.
A particle is considered to be in the signal hemisphere (that which is expected to have the least total energy) if $\cos(\theta)>0$ and in the non-signal hemisphere if $\cos(\theta)<0$.

Signal candidates are constructed by requiring two opposite sign tracks originating from the same position and displaced from the primary interaction. A mass window cut, described in Table \ref{tab:candidates}, is applied to the intermediate $Y$ resonance.

\begin{table}[ht!]
	\centering
	\begin{tabular}{@{}l c c c@{}}
        \toprule
        Decay & Candidate & Candidate Children & Candidate Mass Range [GeV] \\
        \midrule
        \BdKstNuNu & \Kstarz & \Kpm\pimp & [0.65, 1.10] \\
        \BsPhiNuNu & $\phi$ & \Kp\Km & [1.00, 1.06] \\
        \bottomrule
	\end{tabular}
	\caption{\label{tab:candidates}%
        The children PID and candidate mass range required for constructing the candidate particle for each signal decay.
    }
\end{table}

Events are required to have at least one primary vertex, have at least one intermediate $Y$ candidate and the momentum of the intermediate candidate must point towards the minimum energy hemisphere, \ie the candidate must have $\cos(\theta)>0$.

We train two different BDTs to isolate signal candidates from the background.
The first is designed to select based on the overall event topology and energy distribution. 
The second is designed to select based on specific information related to the intermediate candidate.
The \xgboost package~\cite{xgb} is used to train the BDTs using the $k$-fold cross validation method (with $k=4$) to avoid over-training and re-use of events.
Separate trainings are performed for the \BdKstNuNu and \BsPhiNuNu modes, with dedicated signal samples. 
The background training sample uses inclusive samples of \Zbb, \Zcc and \Zqq (with $q\in\{\uquark, \dquark, \squark\})$ appropriately weighted according to the known hadronic \Z branching fractions: 0.1512 (\Zbb), 0.1203 (\Zcc) and $0.4276$ (\Zqq)~\cite{PDG2022}.

\subsection{First-stage BDT}

The first stage BDT is trained using a sample of 1 million signal events and 1 million background events.
The BDT is trained using the following input variables:
\begin{itemize}
    \item The total reconstructed energy in each hemisphere,
    \item The total charged and neutral reconstructed energies of each hemisphere,
    \item The charged and neutral particle multiplicities in each hemisphere,
    \item The number of charged tracks used in the reconstruction of the primary vertex,
    \item The number of reconstructed vertices in the event,
    \item The number of candidates in the event
    \item The number of reconstructed vertices in each hemisphere,
    \item The minimum, maximum and average radial distance of all decay vertices from the primary vertex.
\end{itemize}

\begin{figure}
    \centering
    \includegraphics[width=0.48\textwidth]{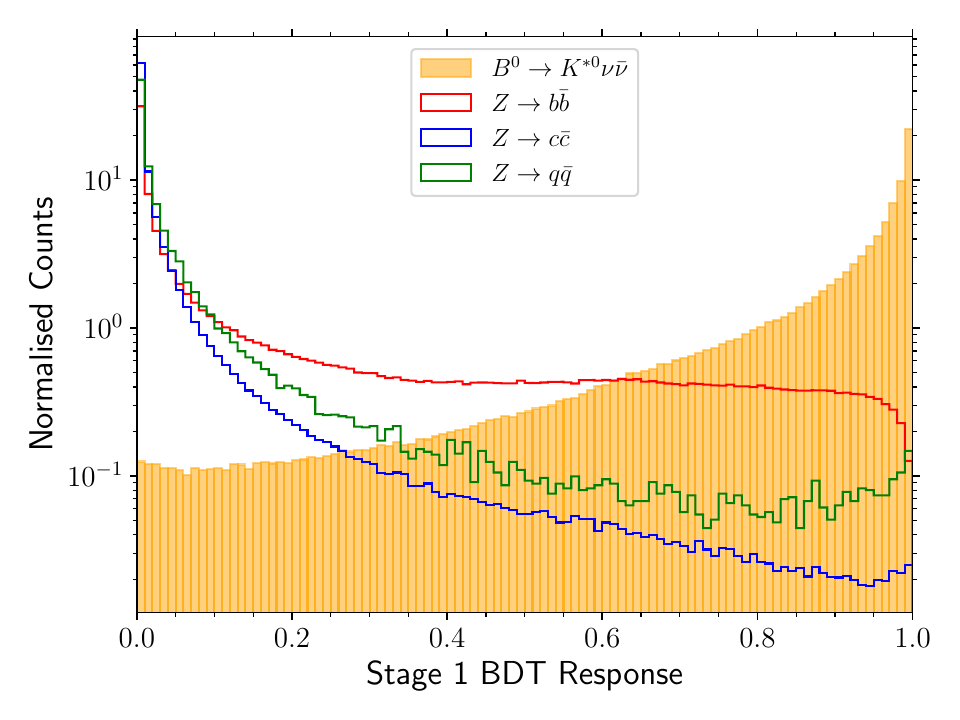} \includegraphics[width=0.48\textwidth]{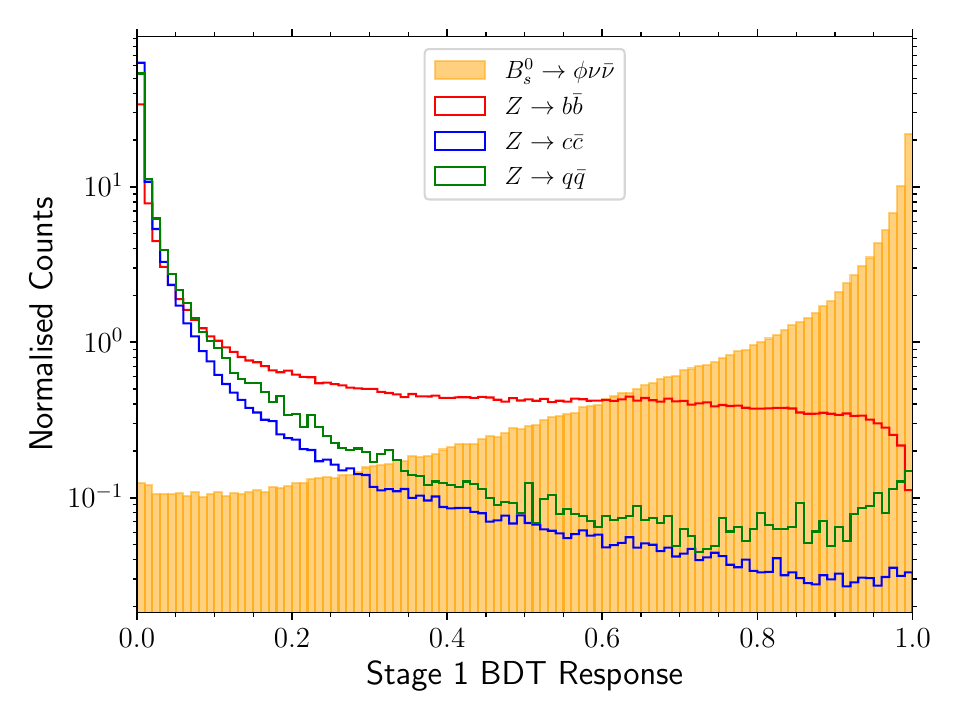}
    \caption{First stage BDT response for the \BdKstNuNu channel (left) and the \BsPhiNuNu channel (right). The relevant signal mode response is shown as the orange filled histogram, the inclusive background sample responses are shown in red, blue and green for \Zbb, \Zcc and \Zqq (for $q\in \{ \uquark, \dquark, \squark\}$), respectively.}
    \label{fig:bdt_st1}
\end{figure}

\begin{figure}
    \centering
    \includegraphics[width=0.48\textwidth]{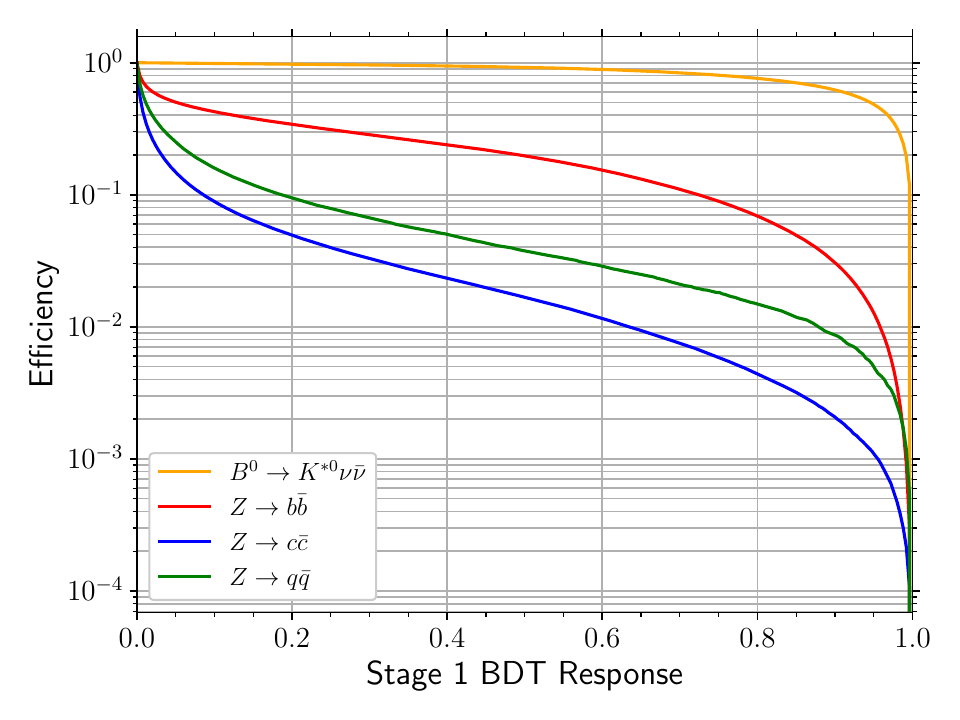} \includegraphics[width=0.48\textwidth]{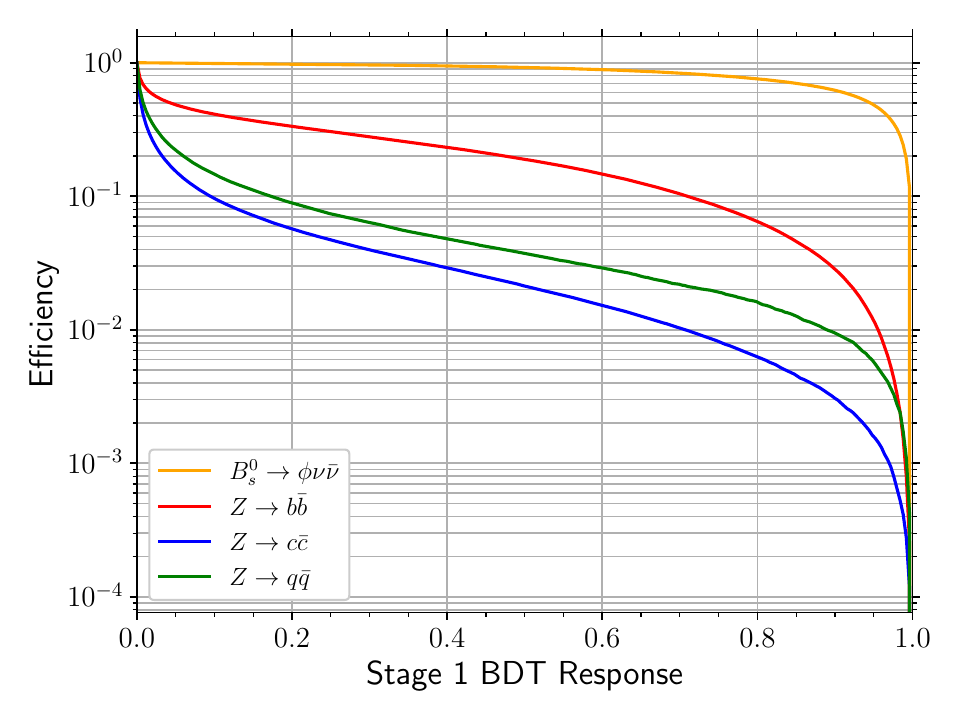}
    \caption{First stage BDT response cut efficiencies for the \BdKstNuNu channel (left) and the \BsPhiNuNu channel (right). The relevant signal mode response is shown as the orange line, the inclusive background sample responses are shown in red, blue and green for \Zbb, \Zcc and \Zqq (for $q\in \{ \uquark, \dquark, \squark\}$), respectively. }
    \label{fig:bdt_st1_eff}
\end{figure}

Figure~\ref{fig:bdt_st1} shows the BDT response in each of the reconstructed channels and Fig.~\ref{fig:bdt_st1_eff} shows the efficiency as a function of a cut on the minimum BDT response. 
It can be seen that the stage 1 BDT is effective at rejecting the inclusive backgrounds, particularly from the lighter quark species, although there is a small mis-identification rate at high BDT scores. 
The integrated ROC score is 0.965 for both the \BdKstNuNu and \BsPhiNuNu channels.

\subsection{Detailed study of background contributions}

\begin{figure}
    \centering
    \includegraphics[width=0.48\textwidth]{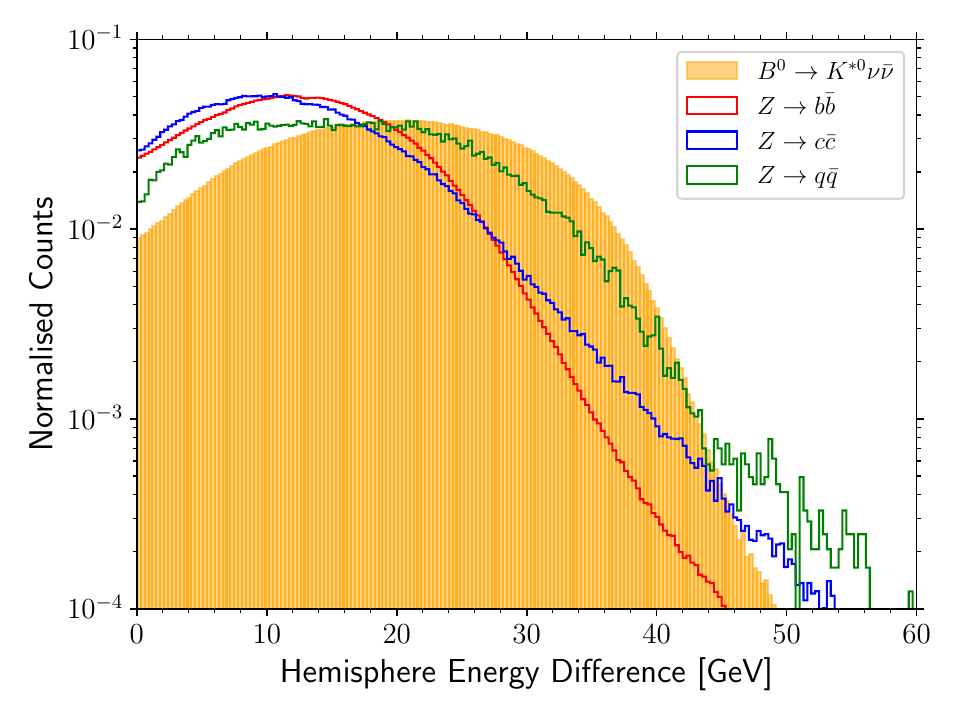}
    \includegraphics[width=0.48\textwidth]{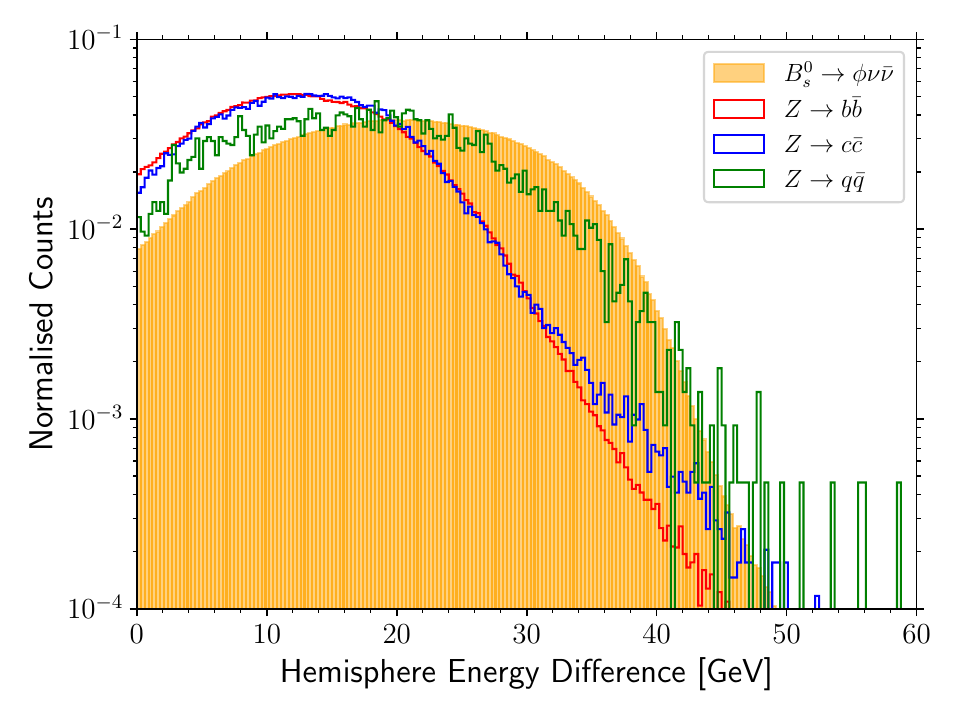}
    \caption{Distributions of the energy difference between the two hemispheres, after a loose cut on $\mathrm{BDT1}>0.6$, for the \BdKstNuNu channel (left) and \BsPhiNuNu channel (right). The relevant signal mode response is shown as the orange filled histogram, the inclusive background sample responses are shown in red, blue and green for \Zbb, \Zcc and \Zqq (for $q\in \{ \uquark, \dquark, \squark\}$), respectively.}
    \label{fig:ediff}
\end{figure}

\begin{figure}
    \centering
    \includegraphics[width=0.48\textwidth]{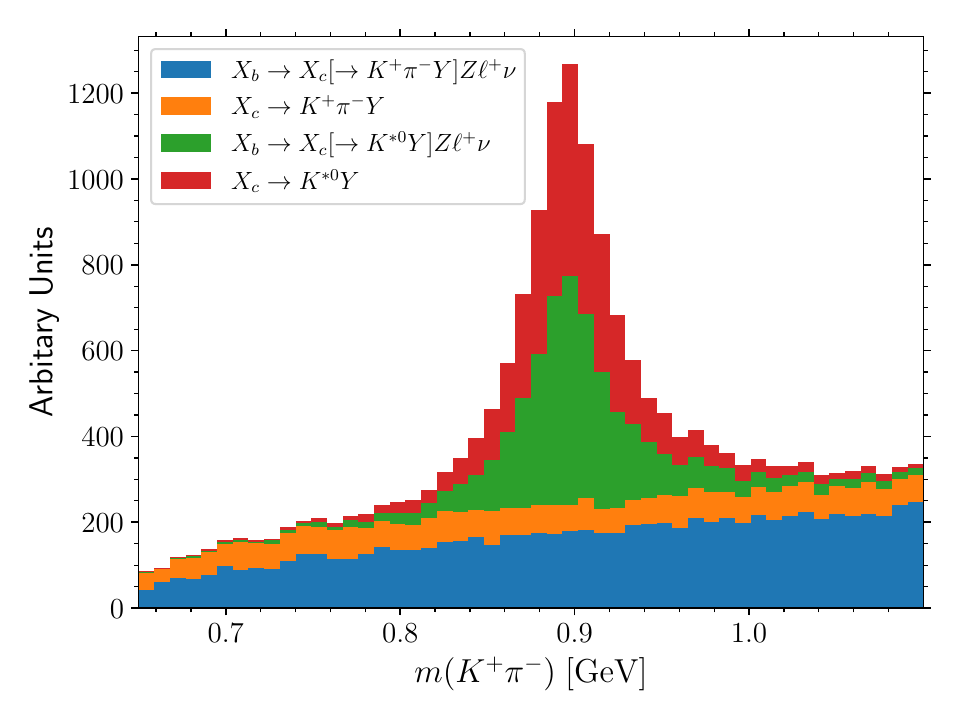}
    \includegraphics[width=0.48\textwidth]{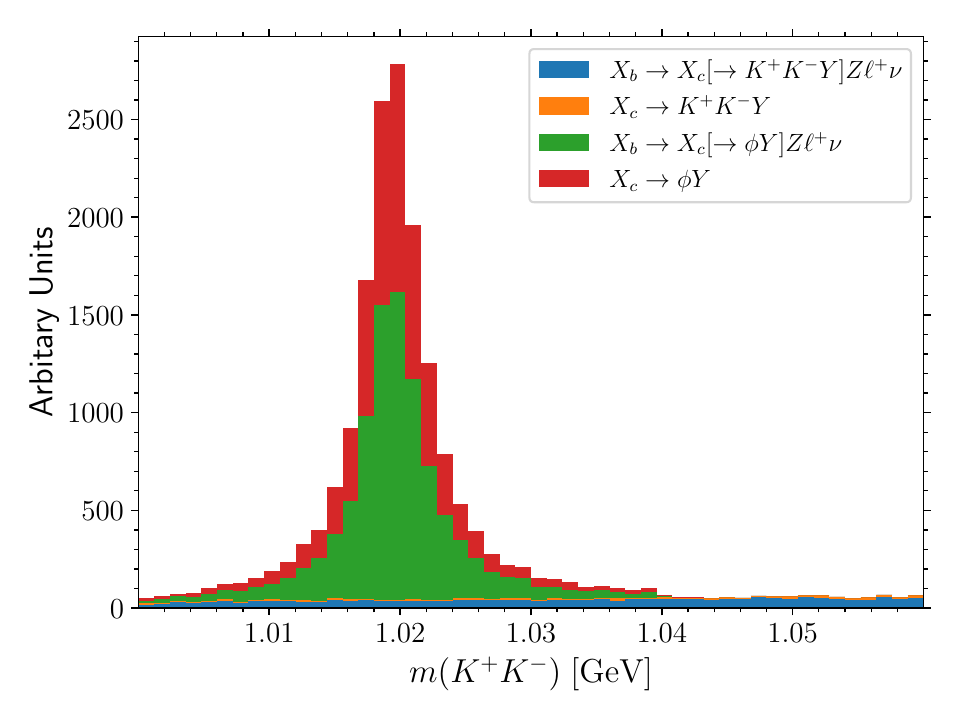}
    \caption{Background contributions as a function of the intermediate resonance mass, for inclusive background events which pass the loose pre-selection, in the \BdKstNuNu mode (left) and \BsPhiNuNu mode (right).
    Contributions are summed over \Zbb, \Zcc and \Zqq with appropriate weighting for their relative branching fractions and selection efficiencies. Each distribution contains two non-resonant components (S-wave) in blue and orange, and two resonant components (left: vector $K^{*}(892)^0$, right: vector $\phi(1020)$) in green and red. The two further distinctions are made between decays originating from a \bquark-hadron (blue and green), labelled $X_b$, and those originating from a prompt \cquark-hadron (orange and red), labelled $X_c$. All of the dominant backgrounds originating from a \bquark-hadron proceed via a secondary \cquark-hadron.}
    \label{fig:bdt_st1_bkgs}
\end{figure}

After the stage 1 BDT we introduce some loose pre-selection cuts which remove a large fraction of the inclusive backgrounds. 
These cuts are on the energy difference between the two hemispheres, $E_{\text{diff}} > 5\gev$, and on the stage 1 BDT, $\mathrm{BDT1} > 0.6$. 
The stage 1 BDT efficiency, shown in Fig.~\ref{fig:bdt_st1_eff}, demonstrates that the cut of $\mathrm{BDT1} > 0.6$ retains $\sim 95\%$ of the signal whilst rejecting $\sim 90\%$ of the inclusive background. 
The distribution of $E_{\text{diff}}$ is shown in Fig.~\ref{fig:ediff}, after the loose cut of $\mathrm{BDT1} > 0.6$ is applied.
By studying in detail, via use of matching to the true MC candidates, the contributions from events which pass these loose cuts, we investigate what sort of backgrounds would be largest in a real-life study.
The results are shown in Fig.~\ref{fig:bdt_st1_bkgs}. The dominant backgrounds are those which proceed via semi-leptonic $\bquark \to \cquark \to \squark$ transitions and semi-leptonic prompt $\cquark\to\squark$ transitions. The most problematic of these are those that contain either real resonant $K^*(892)^0$ or $\phi(1020)^0$, which peak in the relevant invariant mass. The most substantial backgrounds in both decay modes are from $\Bp \to \Dstarz \ellp \neu$ where $\Dstarz\to\Dz\piz$ or $\Dstar\to\Dz\g$ and then the $\Dz$ produces either the \Kstarz or $\phi$ candidate.
A more detailed list of the specific exclusive background modes which contribute most significantly, along with their expected rates, are provided in Appendix~\ref{app:backgrounds}.
These specific backgrounds are not further studied in this work, although they are included as part of the inclusive samples we use to model our background.
The most dominant contributions would require dedicated treatment for future works aiming to maximise the sensitivity.

\subsection{Second-stage BDT}

\begin{figure}
    \centering
    \includegraphics[width=0.48\textwidth]{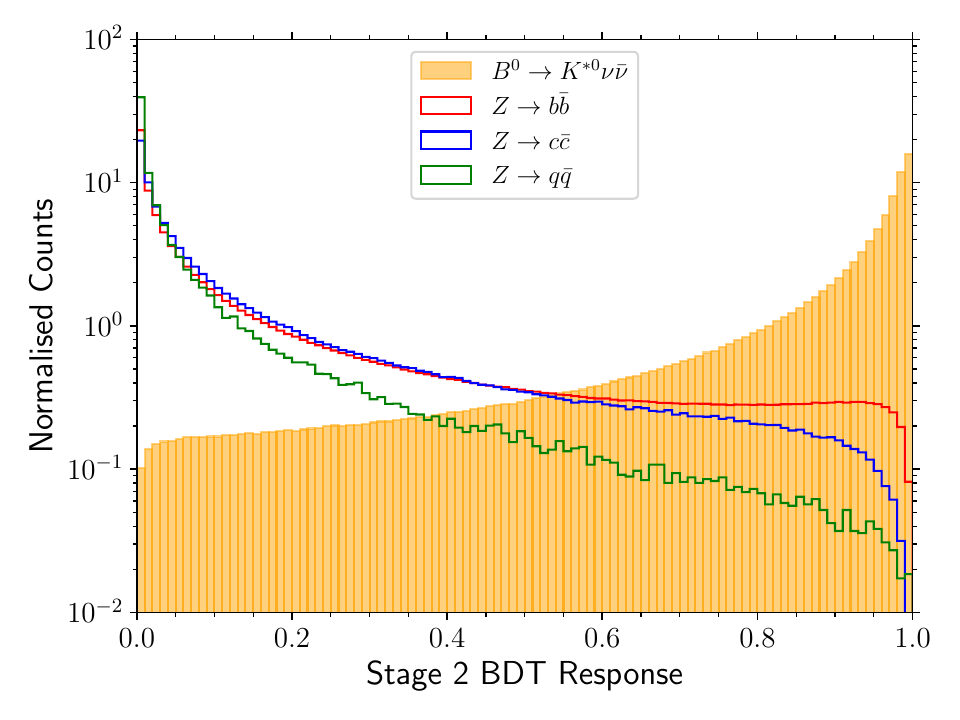} \includegraphics[width=0.48\textwidth]{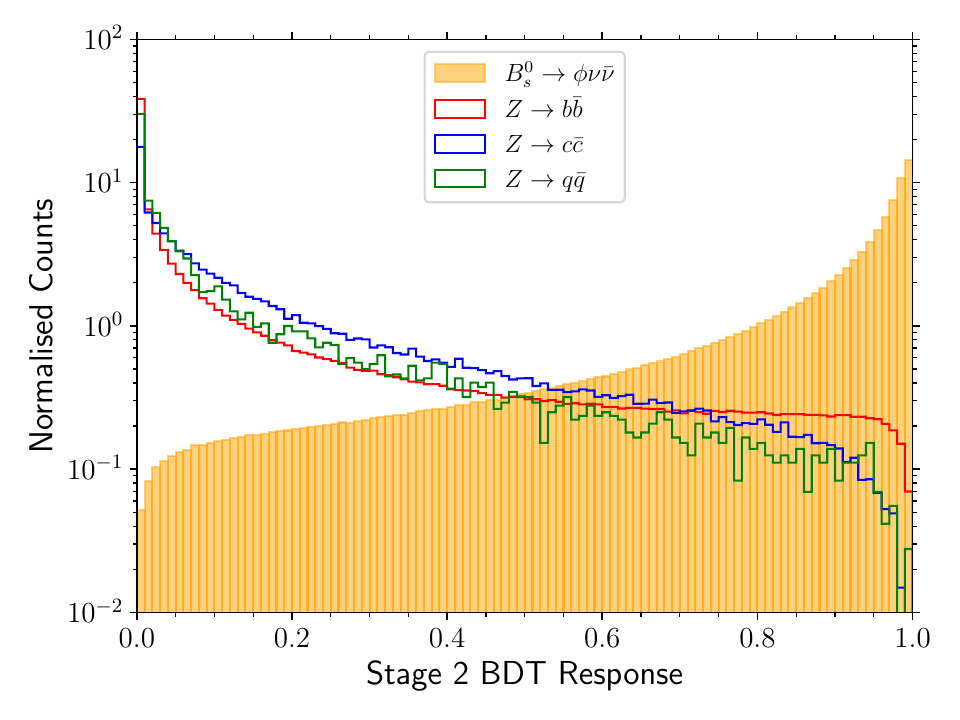}
    \caption{Second stage BDT response for the \BdKstNuNu channel (left) and the \BsPhiNuNu channel (right). The relevant signal mode response is shown as the orange filled histogram, the inclusive background sample responses are shown in red, blue and green for \Zbb, \Zcc and \Zqq (for $q\in \{ \uquark, \dquark, \squark\}$), respectively. }
    \label{fig:bdt_st2}
\end{figure}

The second-stage BDT is trained using a sample of 1 million signal events and 1 million background events which pass the preselection criteria of $E_{\text{diff}}>5\gev$ and $\mathrm{BDT1} > 0.6$.
The second-stage BDT is trained using the following input variables: 
\begin{itemize}
    \item The intermediate candidate's reconstructed mass
    \item The number of intermediate candidates in the event
    \item The intermediate candidate's flight distance and flight distance \chisq from the primary vertex 
    \item The $x$, $y$ and $z$ components of the intermediate candidate's momentum
    \item The scalar momentum of the intermediate candidate
    \item The transverse and longitudinal impact parameter of the intermediate candidate
    \item The minimum, maximum and average transverse and longitudinal impact parameters of all other reconstructed decay vertices in the event
    \item The angle between the intermediate candidate and the thrust axis
    \item The mass of the primary vertex
    \item The nominal $B$ candidate energy, defined as the \Z mass minus all of the reconstructed energy apart from the candidate children
\end{itemize}
Figure~\ref{fig:bdt_st2} shows the second-stage BDT response in each of the reconstructed channels.
The integrated ROC scores are 0.961 and 0.959 for the \BdKstNuNu and \BsPhiNuNu channels, respectively. 

\subsection{Sensitivity Estimate}

In order to obtain an estimate of the overall sensitivity we need to find an optimal cut point in both BDT scores given a particular value of the expected $B\to Y\neu\neub$ branching fraction, where $Y$ is the intermediate resonance, $Y \in \{\Kstarz,\phi\}$.
Given that a combination of cuts on both BDTs is incredibly efficient at rejecting the background we cannot get an accurate estimate of the cut efficiencies directly from the inclusive background samples, because so little of the MC statistics remain for the inclusive backgrounds at high BDT cut values.
Consequently we build a map of the signal and inclusive background efficiencies, $\epsilon^{s}$ and $\epsilon^{b}$, as a function of the two BDT score cut values and then use a bi-cubic spline to interpolate between points.
We then define a figure of merit (FOM) defined as,
\begin{equation}
    \mathrm{FOM} = \frac{S}{\sqrt{S+B}},
\end{equation}
where $S$ is the expected number of signal events and $B$ is the expected number of background events based on the sum of contributions from \Zbb, \Zcc and \Zqq (for $q\in\{u,d,s\})$.

The signal expectation is computed as,
\begin{equation}
    \label{eq:NS}
    S = 2 \, N_Z \,  \BF(\Z\to\bbbar) \, f_B \, \BF(\B\to Y\neu\neub) \, \BF (Y\to f) \, \epsilon^{s}_{\text{pre}} \, \epsilon^{s}_{\text{BDTs}},
\end{equation}
where $N_Z$ is the number of $Z$ bosons produced, the factor of two accounts for the fact there are two \bquark-quarks, $f_B$ is the production fraction for the \bquark-quark to hadronise into the relevant \bquark-hadron, $\BF(\B\to Y\neu\neub)$ is the predicted branching fraction for the decay of interest, $\BF(Y\to f)$ is the branching fraction of the intermediate resonance to the final state $f$, $\epsilon^{s}_{\text{pre}}$ is the signal efficiency of the pre-selection (including the reconstruction and the loose cut on $\mathrm{BDT1}$), and $\epsilon^{s}_{\text{BDTs}}$ is the signal efficiency of the two BDT score cuts.

The background expectation is computed as,
\begin{equation}
    \label{eq:NB}
    B = \displaystyle\sum_{f \in \{\bbbar, \ccbar, \qqbar\}} N_Z \, \BF(\Z \to f) \, \epsilon^{b}_{f,\text{pre}} \, \epsilon^{b}_{f,\text{BDTs}},
\end{equation}
where $\BF(\Z\to f)$ are the relevant branching fractions for $Z\to\text{hadrons}$ (either \bbbar, \ccbar or \qqbar) and $\epsilon^{b}_{f,\text{pre}}$, $\epsilon^{b}_{f,\text{BDTs}}$ are the pre-selection and BDT cut efficiencies of the relevant background, respectively.

\begin{figure}
    \centering
    \includegraphics[width=0.48\textwidth]{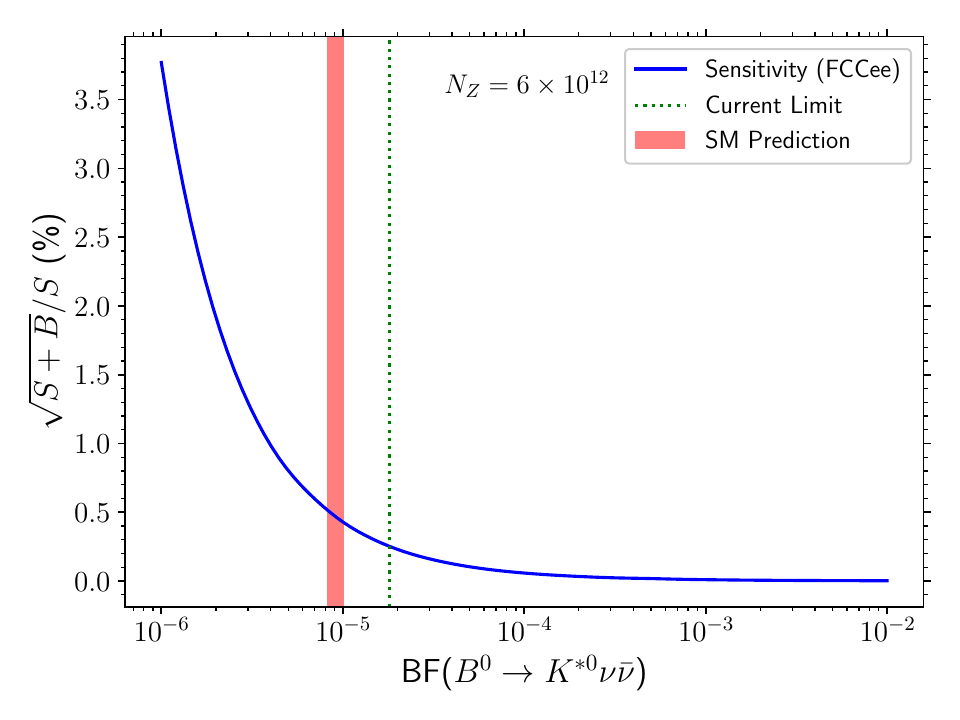}
    \includegraphics[width=0.48\textwidth]{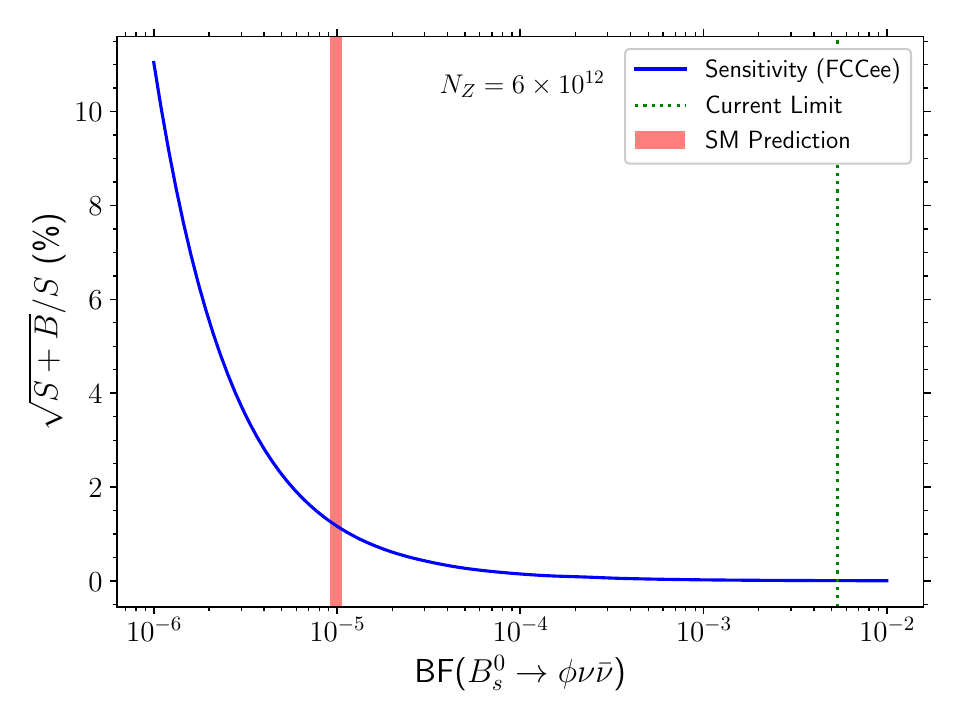}
    \caption{The expected sensitivity of the Tera-$Z$ run at FCC-ee to the branching fraction of \BdKstNuNu (left) and \BsPhiNuNu (right).}
    \label{fig:sensitivity}
\end{figure}

For our study we assume the following values of the parameters in Eqs.~(\ref{eq:NS}) and~(\ref{eq:NB}):
\begin{itemize}
    \item $N_Z=6\times 10^{12}$, the number of \Z-bosons produced across all experiments during the entire Tera-\Z run at FCC-ee.
    \item The production fraction of $\B$-mesons from \Zbb decays are $f_\Bd=0.43$ and $f_\Bs=0.096$.
    \item The SM predictions of the relevant decay branching fractions are provided above in Table~\ref{tab:SMpred}, although we also scan the sensitivity as a function of these branching fractions below (see Fig.~\ref{fig:sensitivity}).
    \item The intermediate resonance branching fractions are $\BF(\Kstarz\to\Kp\pim)=0.9975$ and $\BF(\phi\to\Kp\Km)=0.491$.
    \item The $\Z\to\text{hadrons}$ branching fractions are $\BF(\Zbb) = 0.1512$, $\BF(\Zcc)=0.1203$ and $\BF(\Zqq)=0.4276$.
\end{itemize}

The optimal BDT cuts are then determined by maximising the value of the FOM with respect to the BDT cut values.
The sensitivity provided in units of $\sqrt{S+B}/S$ (\%), in other words the expected relative size of the 1$\sigma$ uncertainty on the measured branching fraction as a function of the hypothesised branching fraction, is shown in Fig.~\ref{fig:sensitivity}.
At the SM predictions the expected sensitivities are 0.53\% for \BdKstNuNu and 1.20\% for \BsPhiNuNu.
The expected number of signal and background events, along with our analysis chain efficiencies are shown in Table~\ref{tab:results}.
A comparison between generated and selected signal candidates in various kinematic distributions is provided in Appendix~\ref{app:kinematic_distributions}.

Given the excellent expected precision to the branching fractions, it would also be feasible to fit the differential branching fractions as a function of $q^2$, therefore allowing for direct measurements of $F_L$. 
Due to the fact that the two neutrinos are not detected, the $q^2$ cannot be measured directly. 
However, using the known beam energy, $b$-hadron momentum and visible energy in both hemispheres, an approximation of the $q^2$ can be iteratively computed using the method described in Ref.~\cite{Li:2022tov}. This is expected to provide a measure of the $q^2$ with a resolution of $\mathcal{O}(2 \gev^2)$.
Based on projections made by the Belle II collaboration for prospects in \bsnunu decays~\cite{Belle-II:2018jsg} we expect that $F_L$ could be measured with a relative uncertainty of $\sim 2.5\%$ in the \BdKstNuNu mode and $\sim 5\%$ in the \BsPhiNuNu mode at FCC-ee.

\begin{table}
    \centering
    \renewcommand{\arraystretch}{1.2}
    \begin{tabular}{l|c c c c c c c c }
         Mode & $N_S$ & $N_B$ & $\epsilon^s$ & $\epsilon^{\bbbar}$ & $\epsilon^{\ccbar}$ & $\epsilon^{\qqbar}$ & $S/B$ & $\sqrt{S+B}/S$ \\
         \hline
         \BdKstNuNu & $231\,$K & $1.27\,$M & 3.7\% & $\mathcal{O}(10^{-7})$ & $\mathcal{O}(10^{-9})$ & $\mathcal{O}(10^{-9})$ & 0.17 & 0.53\% \\
         \BsPhiNuNu & $61\,$K & $0.48\,$M & 7.4\% & $\mathcal{O}(10^{-7})$ & $\mathcal{O}(10^{-9})$ & $\mathcal{O}(10^{-9})$ & 0.13 & 1.20\%
    \end{tabular}
    \caption{The expected number of signal and background events of the Tera-$Z$ run at FCC-ee at the SM predictions for the \BdKstNuNu and \BsPhiNuNu decays. The signal and inclusive background efficiencies of our analysis chain are also shown along with the signal-to-background ratio and the expected sensitivity.}
    \label{tab:results}
\end{table}

\subsection{Extrapolation to neutral modes}

Recent studies of neutral reconstruction performance with IDEA at FCC-ee suggest that the \KS and \Lz reconstruction efficiency is $\sim 80\%$ in the momentum range relevant for this analysis~\cite{Aleksan:2021fbx}. 
Based on the typical efficiencies of our analysis in the \BdKstNuNu and \BsPhiNuNu decays, along with an additional $80\%$ reconstruction efficiency for the \BdKSNuNu and \LbLzNuNu modes, we extrapolate our sensitivity estimates for the neutral modes using Eqs.~(\ref{eq:NS}) and (\ref{eq:NB}), assuming the same background rejection rate can be achieved.

\begin{figure}
    \centering
    \includegraphics[width=0.48\textwidth]{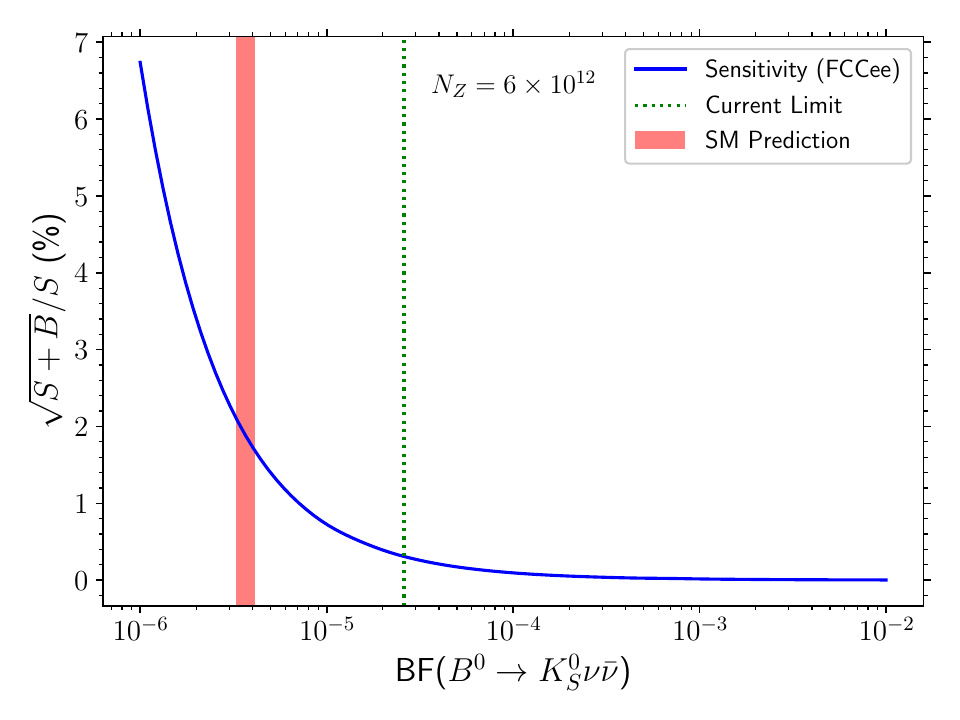}
    \includegraphics[width=0.48\textwidth]{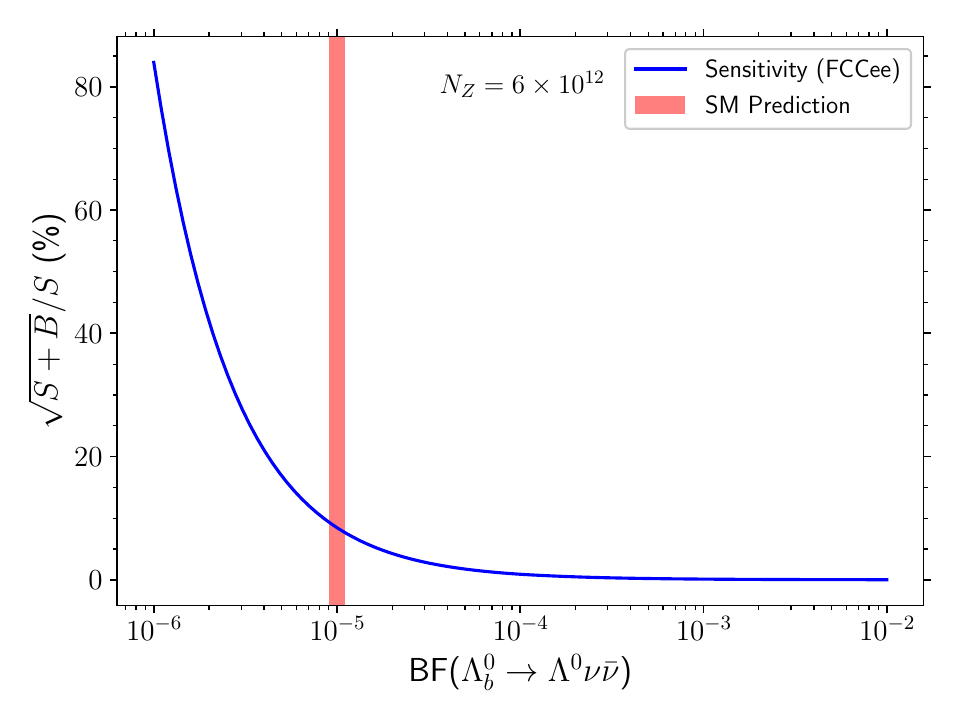}
    \caption{The expected sensitivity of the Tera-$Z$ run at FCC-ee to the branching fraction of \BdKSNuNu (left) and \LbLzNuNu (right).}
    \label{fig:sensitivity_neutral}
\end{figure}

The numerical values used for the terms in Eqs.~(\ref{eq:NS}) and (\ref{eq:NB}) are $f_\Lb = 0.037$, $\BF(\KS\to\pip\pim)=0.692$ and $\BF(\Lz\to \proton\pim)=0.639$.
This results in expected sensitivities (signal-to-background ratios), at the SM prediction, of 3.37\% (0.04) for \BdKSNuNu and 9.86\% (0.015) for \LbLzNuNu.
The extrapolated sensitivity as a function of the hypothesised branching fraction for these modes is shown in Fig.~\ref{fig:sensitivity_neutral}.

\subsection{Study of particle-identification}

As mentioned above, the sensitivity estimates provided in Fig.~\ref{fig:sensitivity} are based on the assumption of perfect particle-identification performance. 
In other words it is assumed that all pions and kaons can be perfectly distinguished by the detector and are thus given the correct mass hypothesis.
This assumption is checked by recomputing the signal efficiencies, $\epsilon^s_{\text{pre}}$ and $\epsilon^s_{\text{BDT}}$ of Eq.~(\ref{eq:NS}), after making random mass hypothesis swaps of kaon \to pion and pion \to kaon, based on an assumed mis-identification rate, $f_{\text{misid}}$.
This incorporates the effect of double mis-identifications and in most cases will cause events to fall outside of the mass window for the intermediate resonance, listed in Table~\ref{tab:candidates}.

\begin{figure}
    \centering
    \includegraphics[width=0.48\textwidth]{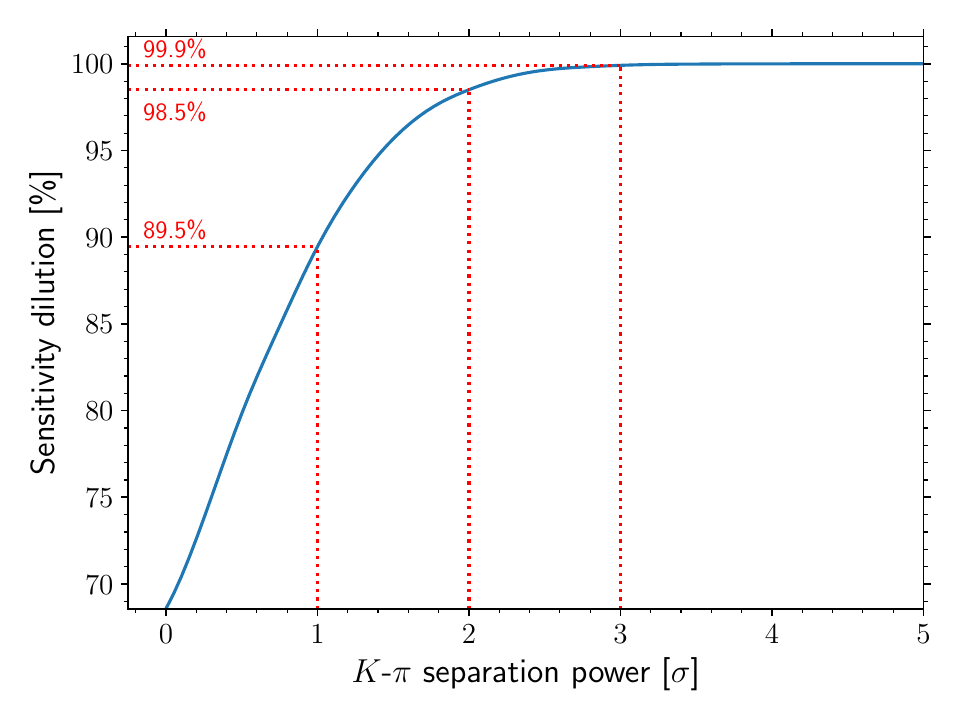}
    \includegraphics[width=0.48\textwidth]{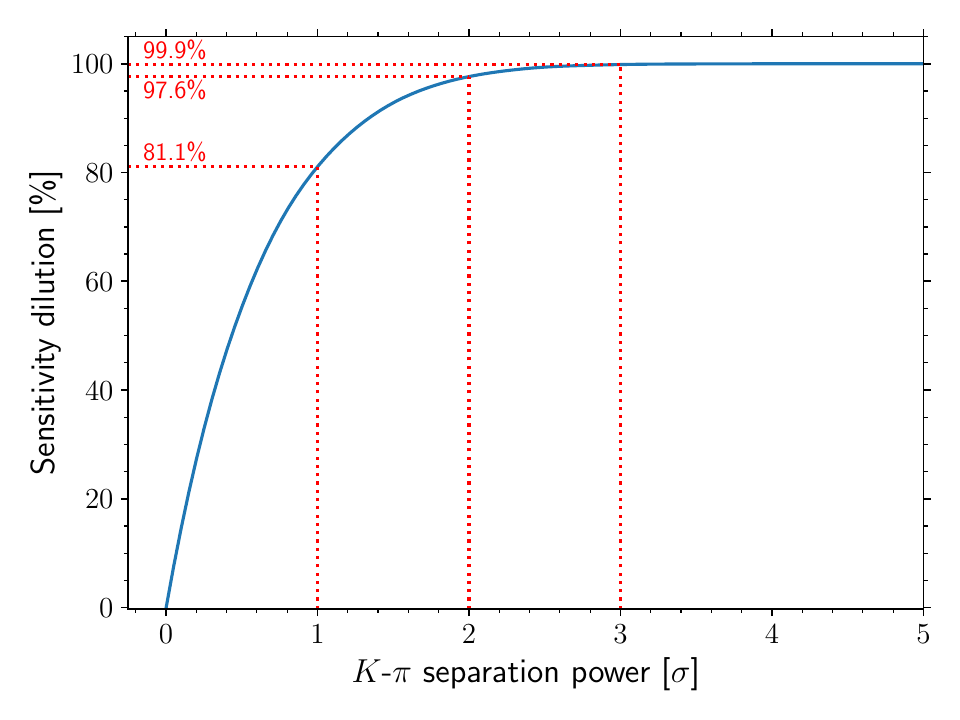}
    \caption{Degradation of the sensitivity to the branching fraction, with respect to the nominal sensitivity assuming perfect PID, as a function of the kaon-pion separation power for the \BdKstNuNu decay (left) and \BsPhiNuNu decay (right).}
    \label{fig:misid}
\end{figure}

The results of this study are shown in Fig.~\ref{fig:misid} in terms of the kaon-pion separation power in standard deviations, $\sigma$, \vs the expected degradation to the sensitivity. 
These show that $K-\pi$ separation of $\sim 2\sigma$ would have a negligible impact on the uncertainty, although the performance rapidly degrades with worse separation.

\subsection{Study of imperfect vertex seeding}

\begin{figure}
    \centering
    \includegraphics[width=0.48\textwidth]{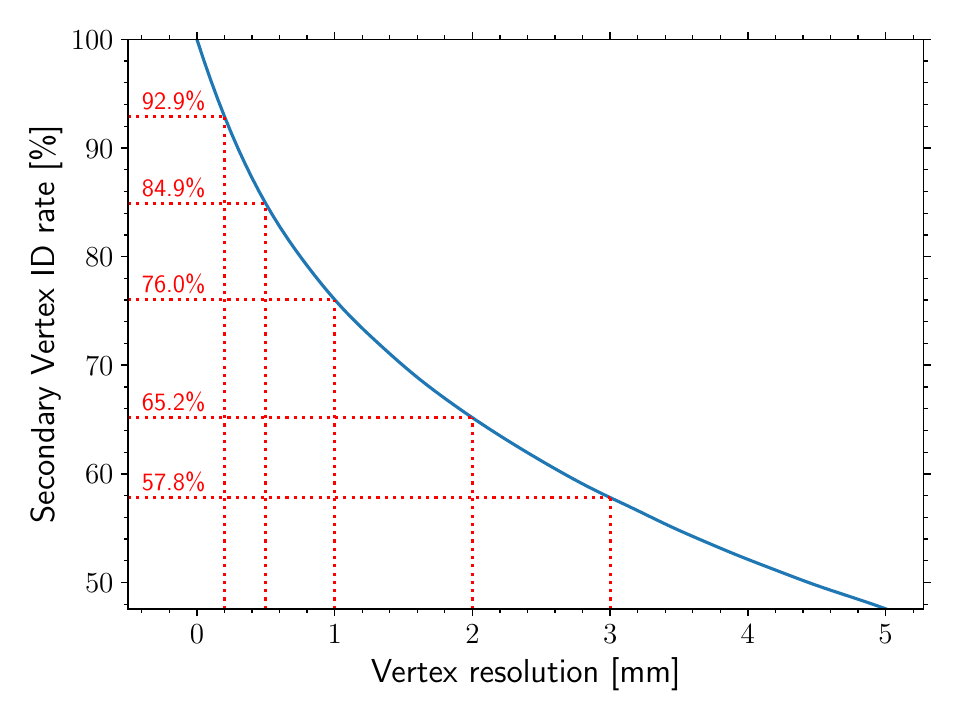}
    \includegraphics[width=0.48\textwidth]{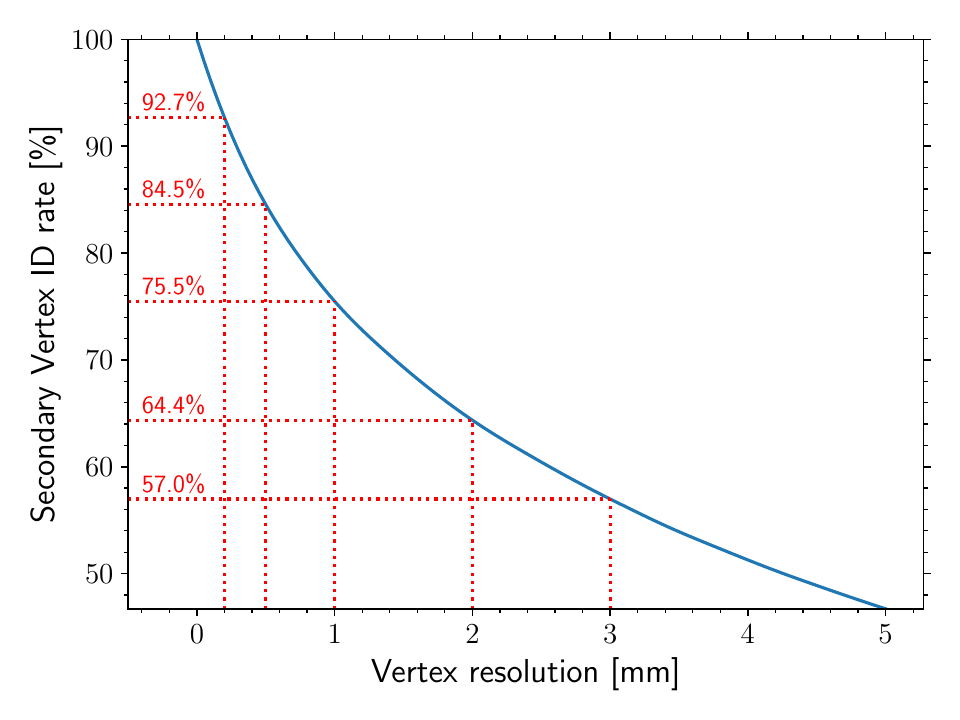}
    \caption{The correct secondary vertex association rate as a function of the expected vertex resolution for the \BdKstNuNu decay (left) and \BsPhiNuNu decay (right).}
    \label{fig:vertex_id}
\end{figure}

Furthermore, the sensitivity estimates provided in Fig.~\ref{fig:sensitivity} assume perfect vertex seeding. 
Whilst the vertex resolution of the detector itself is incorporated it is still assumed that each vertex is correctly identified. 
In practise this will not always be the case and for poorly resolved vertices and for vertices in close proximity it may be that the wrong vertex will be chosen instead.
This effect is investigated by randomly selecting the wrong vertex, based on a value of the vertex resolution, and propagating its effect through the analysis pipeline. 
The results are shown in Fig.~\ref{fig:vertex_id}, which gives the secondary vertex identification rate as a function of the vertex resolution.
This shows that the vertex resolution will need to be $\lesssim 0.2\mm$ in order to sufficiently mitigate vertex mis-identification.
However, this is far above the resolution requirements for vertex precision anyway, $\mathcal{O}(10\mum)$. 
Consequently, we do not expect any significant effect from vertex mis-association.

\subsection{Potential systematic effects}

Our study is somewhat simplistic in that it assumes a pure counting experiment of the signal and background rates above some set of optimised BDT cuts. 
In a real-life study it is likely the sensitivity could be enhanced by fitting the BDT distributions themselves, or indeed by fitting the invariant mass of the intermediate candidate, or by a variety of other as yet unconsidered enhancements.
However, a real-life study will also incur a variety of systematic effects that will need to be considered. 

In terms of the analysis itself, the MC tuning and sample size will be important considerations, as will detailed studies of the most dominant background contributions and detector effects.
In terms of the calculation of the branching fraction from the signal yield, Eq.~\eqref{eq:NS}, there will be additional sources of systematic related to knowledge of the selection efficiencies, hadronisation fragmentation and production fractions, decay multiplicities, and related branching fractions that will need to be considered. 
Many of these quantities remain best measured by the LEP experiments, although with FCC-ee the precision of these measurements will substantially improve, thus reducing the systematic impact on this analysis.

The most significant systematic impact on this analysis arises from knowledge of the $Z\to\bbbar$ branching fraction and the \bquark-quark fragmentation fractions, $f_B$. 
The former is already known from LEP to 3 per mille precision~\cite{PDG2022} and with likely improvements from FCC-ee measurements will not have a significant impact on the precision of this analysis.
The latter, however, is currently only known to $\sim 2\%$ precision~\cite{PDG2022} meaning a potentially significant systematic impact on this analysis if further enhancements are not performed at FCC-ee. 
It is expected that FCC-ee itself will be able to improve knowledge of the fragmentation fractions by an order of magnitude or more, reducing the systematic impact to the same order as the statistical precision.

\FloatBarrier

\section{Phenomenology}
\label{sec:interpretation}

In this section, we investigate the implications of measurements of the \bsnunu observables with the expected sensitivity obtained in the previous sections, namely $0.53\%$ for \BdKstNuNu, $1.20\%$ for \BsPhiNuNu, $3.37\%$ for \BdKSNuNu, and $9.86\%$ \LbLzNuNu.
We will consider the current uncertainties for the SM predictions quoted in Table~\ref{tab:SMpred}, and we will study the impact of an improvement on these uncertainties by more precise and accurate determinations of the CKM factor $|\lambda_t| = |V_{tb}^{\phantom{*}} V_{ts}^*|$ and the hadronic form factors.

\subsection{SM implications}
As discussed in Sec.~\ref{sec:theory}, the two main sources of uncertainties are the product of CKM matrix elements $|\lambda_t|$ and the form factors.
We first assume that NP effects are absent and study how a precise measurement of \bsnunu could provide us with information about these quantities.

\subsubsection*{Extraction of CKM elements.}
As a first illustration of the potential of these measurements at FCC-ee, we study the precision in extracting $|\lambda_t|^2$ from these decays by using the form factors determined from LQCD.
The most convenient decay for this purpose is \BdKSNuNu, for which only a single form factor is needed and is already predicted with a $\approx 5\%$ precision. 
In this case, we can write,
\begin{equation}
    \label{eq:lambdatfromBToK}
    |\lambda_t| = (39.3 \times 10^{-3}) \bigg{[}\dfrac{\mathcal{B}(\BdKSNuNu)^\mathrm{exp}}{2.02 \times 10^{-6}}\bigg{]}^{1/2}\bigg{[}\dfrac{\kappa_+}{24.8}\bigg{]}^{-1/2}
    \quad \mathrm{where} ~ \kappa_+ = \int dq^2 \rho_+^\KS(q^2).
\end{equation}

\begin{figure}
    \centering\includegraphics[width=0.6\textwidth]{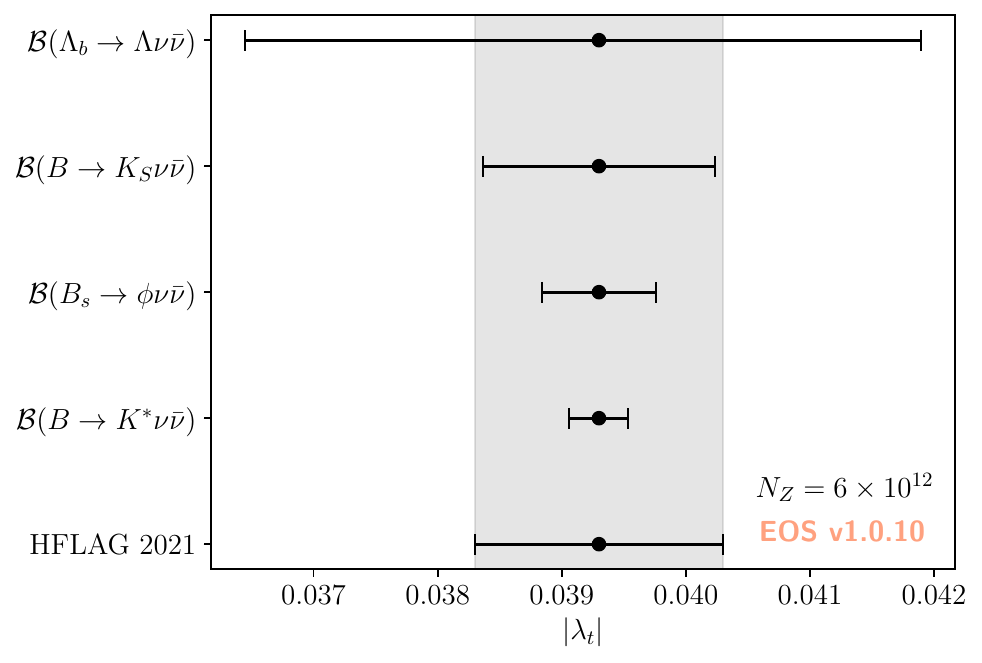}
    \caption{
        68\% probability ranges assuming the branching ratios to be SM-like.
        We used the experimental uncertainties of Sec.~\ref{sec:analysis} and the \textbf{Future} form factors uncertainties described in Sec.~\ref{sec:theory}.
        The results are compared with the value derived from $|V_{cb}|= (40.0 \pm 1.0)\times 10^{-3}$, extracted from $B\to D \ell \nu$ decays~\cite{FlavourLatticeAveragingGroupFLAG:2021npn}.
    }
    \label{fig:ckm_fit}
\end{figure}

Equation~(\ref{eq:lambdatfromBToK}) clearly shows that the joint effort of both the lattice and the experimental communities will make \BdKSNuNu decays a major player in the extraction of $|\lambda_t|$.
We now extend this study to the other modes using the \textbf{Future} form factors uncertainties described in Sec.~\ref{sec:theory}.
The results are shown in Fig.~\ref{fig:ckm_fit}, where the extracted values of $\lambda_t$ are compared to the current world average.

\subsubsection*{Extraction of hadronic form factors.}
Conversely, assuming accurate knowledge of the CKM elements from other sources, the \bsnunu decays allow for a simultaneous extraction of the form factors.
The dependency on CKM elements can also be lifted by considering only the shape of the form factors~\cite{Becirevic:2023aov}.
Assuming the SM, an unnormalised binned likelihood fit of the differential branching ratios provides direct access to the shape of the scalar form factor $\rho_+^K$ and to the combination of the vector form factors $\rho_V + \rho_{A_1} + \rho_{A_{12}}$.

As an example, we extend the definition of the ratio $r_\mathrm{lh}$ introduced in Ref.~\cite{Becirevic:2023aov} to the other modes using
\begin{equation}
    r_\mathrm{lh}^{Y} = \frac{\mathcal{B}(B\to Y \nu \bar{\nu})_{0 < q^2 < q^2_\mathrm{max} / 2}}{\mathcal{B}(B\to Y \nu \bar{\nu})_{q^2_\mathrm{max} / 2 < q^2 < q^2_\mathrm{max}}}.
\end{equation}
Assuming the current uncertainties on the form factors, we predict
\begin{equation}
    r_\mathrm{lh}^{K} = 1.91(6), \quad r_\mathrm{lh}^{K^*} = 0.84(6), \quad r_\mathrm{lh}^{\phi} = 0.96(9), \quad r_\mathrm{lh}^{\Lambda} = 0.50(9).
\end{equation}
With the benchmark \textbf{Future} form factors we get
\begin{equation}
    r_\mathrm{lh}^{K} = 1.91(1), \quad r_\mathrm{lh}^{K^*} = 0.83(1), \quad r_\mathrm{lh}^{\phi} = 0.82(1), \quad r_\mathrm{lh}^{\lambda} = 0.49(1).
\end{equation}
The interest of these ratios is clear from the reduced uncertainty one finds already with the current form factors.
The effect is even more striking when the uncertainty on the form factors is smaller.
This demonstrates that these ratios will provide valuable information when extrapolating the form factors from high $q^2$, where the lattice QCD results are the most precise, to the low $q^2$ region.
This method can eventually be extended, once the statistical power allows it, to a full unnormalised binned likelihood fit to all of the available differential observables.

\subsubsection*{Ratio of charged and neutral leptons.}
Finally, we emphasize the interest of ratios of the form
\begin{equation}
    R_Y^{\ell/\nu} = \frac{\mathcal{B}(B \to Y \ell^+ \ell^-)}{\mathcal{B}(B \to Y \nu\bar{\nu})},
\end{equation}
where $\ell$ is a charged lepton and the branching ratio can be integrated over the full kinematical range or, according to the experimental precision, over several bins.
These ratios benefit from numerous uncertainty cancellations, both from the experimental side (fragmentation fraction, branching fraction of the normalization channel, experimental efficiency \etc) and the theory side (CKM elements, local form factors \etc)~\cite{Becirevic:2023aov}.

Experimentally, $R_{K^+}^{\mu/\nu}$ can be reconstructed using the world average measurement of $B\to K\mu^+\mu^-$ decays~\cite{PDG2022} and the combination of the searches and observation of $B\to K \neu \neub$ presented by the Belle II collaboration~\cite{Belle-II:2023esi}.
Assuming uncorrelated uncertainties, we obtain
\begin{equation}
    R_{K^+}^{\mu/\nu}|_{2023} = 0.03 \pm 0.01.
\end{equation}
For the other modes, only lower limits can be set.
Using again world averages~\cite{PDG2022}, we get at $90\%$~CL
\begin{equation}
    R_{K^{*+}}^{\mu/\nu}|_{2023} > 0.02, \qquad
    R_{K^{*0}}^{\mu/\nu}|_{2023} > 0.07, \qquad
    R_{\phi}^{\mu/\nu}|_{2023} > 2 \times 10^{-4}.
\end{equation}

Reliable theoretical predictions of $R_M^{\ell/\nu}$ are challenged by long-range effects, dominated by the charm loops and subject to several approaches \cite{Ciuchini:2022wbq,Gubernari:2022hxn}.
These effects give rise to a shift to the Wilson coefficient $C_9^\ell$ that enters the Hamiltonian relevant to the $bs\ell\ell$ sector of the WET.
Any measurement of these ratios therefore provides invaluable information for understanding the non-local contributions.
Assuming that the neutrino mode will dominate the experimental uncertainties, the sensitivities expected for FCC-ee will permit a direct extraction of the shift to $C_9^\ell$
with an accuracy of $8.7\%, 13\%, 22\%$ and $37\%$ for the $B\to K$, $B\to K^*$, $B_s\to\phi$ and $\Lb\to\Lz$ transition respectively.

\subsection{NP implications}
Assuming three massless, left-handed neutrino species below the electroweak scale, the dimension-6 effective Hamiltonian in Eq.~(\ref{eq:th:Heff}) is augmented by only one additional contribution from potential NP beyond the SM~\cite{Felkl:2021uxi}
\begin{equation}
    \mathcal{O}_R^{\nu_i,\nu_j} =
        \frac{e^2}{16 \pi^2} \big( \bar{s}_R \gamma_\mu b_L \big) \big( \bar{\nu}_i \gamma^\mu (1-\gamma_5) \nu_j \big).
\end{equation}
Assuming universal flavour conserving contributions only, the $B$-meson observables take the simple form \cite{Felkl:2021uxi}
\begin{align}
    \frac{d\BF(\BdKSNuNu)}{dq^2} & =
        3 \, \tau_{B^0} |N_{\Bd}|^2 |C_L + C_R|^2 |\lambda_t|^2 \rho^\KS_+\,, \\[0.35em]
    \frac{d\BF(\BdKstNuNu)}{dq^2} & =
        3 \, \tau_B |N_{\Bd}|^2 |\lambda_t|^2 \left(|C_L - C_R|^2 (\rho^\Kstarz_{A_1} + \rho^\Kstarz_{A_{12}}) + |C_L + C_R|^2 \rho^\Kstarz_V \right)\,, \\[0.35em]
    \frac{d\BF(\BsPhiNuNu)}{dq^2} & =
        3 \, \tau_{B_s} |N_{\Bs}|^2 |\lambda_t|^2 \left(|C_L - C_R|^2 (\rho^\phi_{A_1} + \rho^\phi_{A_{12}}) + |C_L + C_R|^2 \rho^\phi_V \right)\,,
\end{align}
\begin{align}
    F_L(\BdKstNuNu) &= \frac{|C_L - C_R|^2 \rho^\Kstarz_{A_{12}}}{|C_L - C_R|^2 (\rho^\Kstarz_{A_1} + \rho^\Kstarz_{A_{12}}) + |C_L + C_R|^2 \rho^\Kstarz_V }\,,\\[0.35em]
    F_L(\BsPhiNuNu) &= \frac{|C_L - C_R|^2 \rho^\phi_{A_{12}}}{|C_L - C_R|^2 (\rho^\phi_{A_1} + \rho^\phi_{A_{12}}) + |C_L + C_R|^2 \rho^\phi_V}.
\end{align}
Setting the lepton masses to zero in Ref.~\cite{Boer:2014kda}, we also get
\begin{gather}
    \frac{d\BF(\LbLzNuNu)}{dq^2} =
        3 \, \tau_{\Lb} |N_{\Lb}|^2 |\lambda_t|^2 \left(|C_L - C_R|^2 (\rho^\Lz_{f_\perp^A} + \rho^\Lz_{f_0^A}) + |C_L + C_R|^2 (\rho^\Lz_{f_\perp^V} + \rho^\Lz_{f_0^V})\right), \\[0.35em]
    F_L(\LbLzNuNu) = \frac{|C_L - C_R|^2 \rho^\Lz_{f_0^A} + |C_L + C_R|^2 \rho^\Lz_{f_0^V}}{|C_L - C_R|^2 (\rho^\Lz_{f_\perp^A} + \rho^\Lz_{f_0^A}) + |C_L + C_R|^2 (\rho^\Lz_{f_\perp^V}   + \rho^\Lz_{f_0^V})}, \\[0.35em]
    A_\mathrm{FB}^\Lz(\LbLzNuNu) = \frac{\alpha}{2}
        \frac{\left( |C_L|^2 - |C_R|^2 \right) \left(\tilde{\rho}^\Lz_\perp + \tilde{\rho}^\Lz_0\right)}{|C_L - C_R|^2 (\rho^\Lz_{f_\perp^A} + \rho^\Lz_{f_0^A}) + |C_L + C_R|^2 (\rho^\Lz_{f_\perp^V} + \rho^\Lz_{f_0^V})}.
\end{gather}
Above, we assume that $C_L^{\nu_i,\nu_j} \equiv \delta_{\nu_i\nu_j} C_L$ and $C_R^{\nu_i,\nu_j} \equiv \delta_{\nu_i\nu_j} C_R$.
The full $B$-meson expressions, including flavour violating contributions, can be found in e.g. Ref.~\cite{Felkl:2021uxi}.
The above expressions present a global $(C_L, C_R) \to (-C_L, -C_R)$ symmetry and another symmetry, $(C_L, C_R) \to (C_R, C_L)$, which is only violated by $A_\mathrm{FB}^\Lz(\LbLzNuNu)$.\\

\begin{figure}
    \centering
    \includegraphics[width=0.49\textwidth]{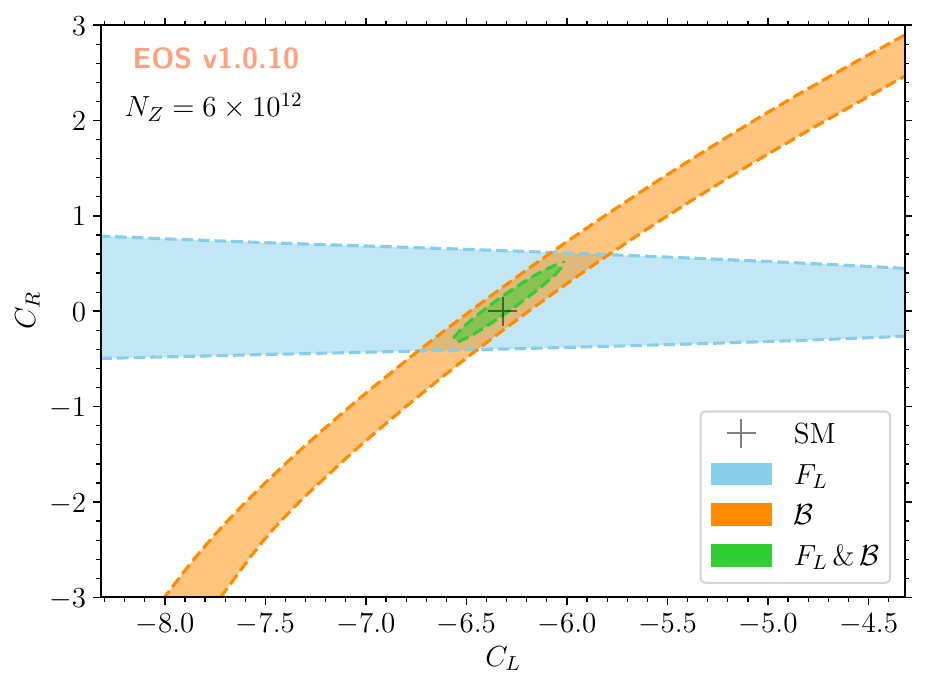}
    \includegraphics[width=0.49\textwidth]{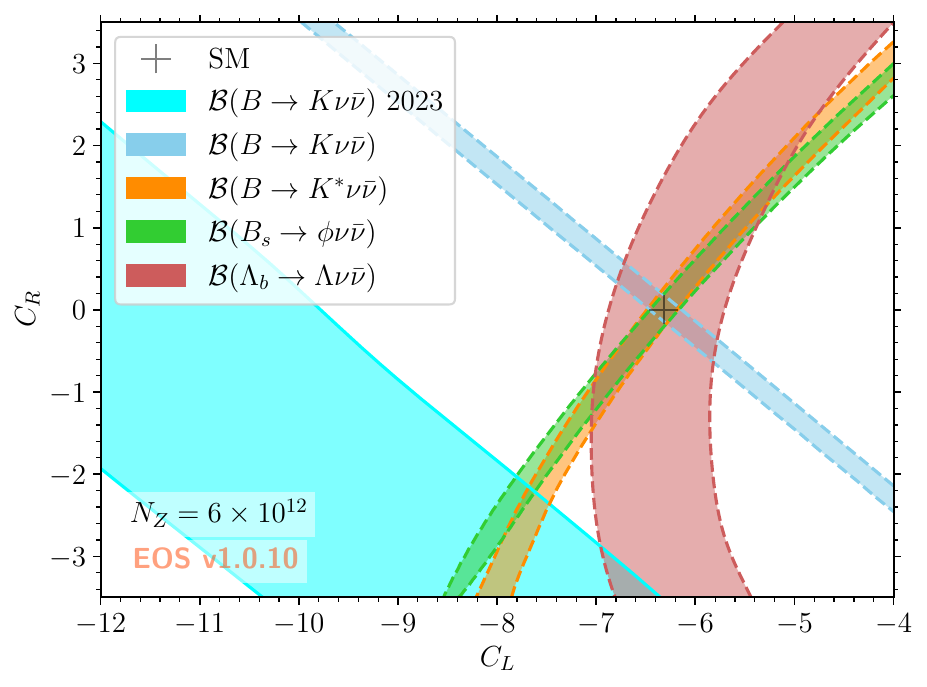}
    \caption{
        Regions with $68\%$ probability of the marginal posterior density, assuming that all observables are SM-like.
        We used the experimental uncertainties of Sec.~\ref{sec:analysis} and the \textbf{Future} form factors uncertainties described in Sec.~\ref{sec:theory}.
        \textbf{Left}: Regions constrained by a measurement of only the branching ratio (orange band), only the longitudinal fraction (blue band) and both (green ellipse) in the case of \BdKstNuNu decays.
        \textbf{Right}: Comparison between the current constraint due to $\mathcal{B}(B\to K\nu\bar{\nu})$ measurements~\cite{Belle-II:2023esi} (cyan band) and the sensitivities predicted at FCC-ee in this study (blue, orange, green and red bands).
    }
    \label{fig:wet_fit}
\end{figure}

The measurement of the \bsnunu branching ratio converts into lines (for $B\to K\nu\bar{\nu}$) or ellipses (for the other channels) in the $(C_L, C_R)$ plane.
On the other hand, longitudinal fractions give cross-shaped constraints.
Up to the 4-fold degeneracy due to the symmetries of these two sets of observables, a BSM point can be unambiguously obtained only by a combined measurement of several branching ratios or longitudinal fractions.
This is depicted in Fig.~\ref{fig:wet_fit}, where the left panel shows the interest of a such a combination in the case of \BdKstNuNu decays.
In the right panel, we compare the estimated constraints obtained at the end of the Tera-Z run to the current constraint derived from the experimental average of $\mathcal{B}(B\to K\nu\bar{\nu})$~\cite{Belle-II:2023esi} (we refer to Refs.~\cite{Athron:2023hmz,Bause:2023mfe,Allwicher:2023syp,Felkl:2023ayn} for more complete WET studies).
We stress that the Belle II measurement is performed assuming SM-like distributions.
Converting this measurement into a constraint on the Wilson coefficients therefore requires a proper reanalysis, which is beyond the scope of this paper.\\

\section{Conclusion}
\label{sec:conclusion}

We carry out an initial performance study on the measurement of \bsnunu decays at FCC-ee running at the $Z$ pole.
To achieve this, we produce updated SM predictions of the observables related to the decays \BdKSNuNu, \BdKstNuNu, \BsPhiNuNu and \LbLzNuNu, both with current and projected theory uncertainties.

We then study the expected sensitivity to these observables, under the assumption that $6\times 10^{12}$ $Z$-bosons are produced in the lifetime of FCC-ee ``Tera-Z" running. 
We find that the uncertainty on the branching fractions, at the SM predicted values, are a relative $0.53\%$, $1.20\%$, $3.37\%$ and $9.86\%$ for the \BdKstNuNu, \BsPhiNuNu, \BdKSNuNu and \LbLzNuNu decays, respectively.
The sensitivity estimates for the neutral \BdKSNuNu and \LbLzNuNu are based on rather simplistic assumptions but a full study of these modes was considered beyond the scope for this paper and will be revisited in future works.

In addition we investigate the impact on the sensitivity of particle identification and vertex identification performance.
For the former we find that the sensitivity is significantly degraded if the kaon-pion separation power is less than $2\sigma$.
For the latter we find no significant impact on vertex seeding providing the vertex resolution is below 0.2~mm.

Finally, we investigate the impact such measurements would have on SM and beyond SM interpretations.
Not only do we find that these decays have a high potential for the extraction of CKM parameters, but we also show that they provide theoretically clean access to the form factors that enter the equivalent decays to charged leptons.
Consideration of the ratios of the branching fraction to the charged and neutral lepton may therefore be the only unambiguous probe of the hadronic effects that plague the interpretation of $b\to s\ell\ell$ decays.

Our studies demonstrate that FCC-ee offers an unparalleled and probably unique opportunity to measure these incredibly rare, experimentally difficult, yet theoretically clean observables with exquisite precision.

\section*{Acknowledgments}

We would like to thank Olcyr Sumensari for his contributions to the early stages of the project as well as Danny van Dyk and Paula Alvarez Cartelle for comments on the manuscript.
We would also like to thank Clement Helsens and Donal Hill for their help with setup, simulation and running of the FCC analysis code.
We thank the FCC-ee Physics Performance Group for the fruitful discussions and helpful feedback on the analysis procedure and manuscript, in particular Guy Wilkinson, Stephane Monteil, Patrizia Azzi, Emmanuel Perez and Xunwu Zuo. 
We also thank our colleagues in the Warwick LHCb group for their helpful advice.
M.R. thanks Stefan Meinel for the discussion on the future of form factor uncertainties.
M.K is supported by the Science and Technology Facilities Council (STFC), UK, under grant \texttt{\#ST/R004536/3} and UK Research and Innovation under grant \texttt{\#EP/X014746/2}. A.R.W. is supported by the STFC, UK.
\clearpage
\appendix

\section{Form factors definition}
\label{app:sec:from-factors}

The three $\bar{B}\to \bar{P}$ form factors are defined by 
\begin{align}
    \langle\bar  P(k) | J_V^\mu | \bar{B}(p) \rangle &=
        \left[ (p + k)^\mu - \frac{M_B^2 - M_P^2}{q^2} q^\mu \right] f_+^{B\to P}
        + \frac{M_B^2 - M_P^2}{q^2} q^\mu f_0^{B\to P}, \\
    \langle\bar  P(k) | J_T^\mu | \bar{B}(p) \rangle &=
        \frac{i f_T^{B\to P}}{M_B + M_P} \left[ q^2 (p + k)^\mu - (M_B^2 - M_P^2) q^\mu \right].
\end{align}
The seven $\bar{B}\to \bar{V}$ form factors are defined by
\begin{align}
    \langle\bar  V(k, \eta) | J_V^\mu |\bar{B}(p) \rangle &=
        \epsilon^{\mu\nu\rho\sigma} \eta_\nu^* p_\rho k_\sigma \frac{2 V^{B\to V}}{M_B + M_V}, \\
    \label{eq:th:param:A}
    \langle\bar  V(k, \eta) | J_A^\mu |\bar{B}(p) \rangle &=
        i \eta_\nu^* \bigg[ g^{\mu\nu} (M_B + M_V) A_1^{B\to V}
        - (p + k)^\mu q^\nu  \frac{A_2^{B\to V}}{M_B + M_V} \nonumber \\
        & \hspace{15mm} - 2 M_V \frac{q^\mu q^\nu}{q^2} (A_3^{B\to V} - A_0^{B\to V}) \bigg],\\
    \langle\bar  V(k, \eta) | J_T^\mu | \bar{B}(p) \rangle &=
        \epsilon^{\mu\nu\rho\sigma} \eta_\nu^* p_\rho k_\sigma \, 2 T_1^{B\to V}, \\
    \langle\bar  V(k, \eta) | J_{AT}^\mu |\bar{B}(p) \rangle &=
        i \eta_\nu^* \bigg[ \Big( g^{\mu\nu} (M_B^2 - M_V^2) - (p + k)^\mu q^\nu \Big) T_2^{B\to V} \nonumber \\
        & \hspace{15mm} - q^\nu \left( q^\mu - \frac{q^2}{M_B^2 - M_V^2} (p + k)^\mu \right) T_3^{B\to V} \bigg],
\end{align}
where $\eta$ is the polarisation four-vector of the vector meson, and we abbreviate
\begin{equation}
    A_3^{B\to V} \equiv \frac{M_B + M_V}{2 \, M_V} \, A_1^{B\to V} - \frac{M_B - M_V}{2 \, M_V} \, A_2^{B\to V}.
\end{equation}
The ten $\bar{\Lambda}_b\to \bar{\Lambda}$ form factors are defined by~\cite{Feldmann:2011xf}
\begin{align}
    \bra{ \Lambda(k,s_\Lambda) } \overline{s} \,\gamma^\mu\, b \ket{ \Lambda_b(p,s_{\Lambda_{b}}) }
        & = \overline{u}_\Lambda(k,s_{\Lambda}) \bigg[ f_t^V(q^2)\: (m_{\Lambda_b}-m_\Lambda)\frac{q^\mu}{q^2} \\
    \nonumber
        &   \phantom{\overline{u}_\Lambda \bigg[}+ f_0^V(q^2) \frac{m_{\Lambda_b}+m_\Lambda}{s_+}
            \left( p^\mu + k^{ \mu} - (m_{\Lambda_b}^2-m_\Lambda^2)\frac{q^\mu}{q^2}  \right) \\
    \nonumber
        &   \phantom{\overline{u}_\Lambda \bigg[}+ f_\perp^V(q^2)
            \left(\gamma^\mu - \frac{2m_\Lambda}{s_+} p^\mu - \frac{2 m_{\Lambda_b}}{s_+} k^{ \mu} \right) \bigg] u_{\Lambda_b}(p,s_{\Lambda_{b}}) \, , \\
    \bra{ \Lambda(k,s_{\Lambda}) } \overline{s} \,\gamma^\mu\gamma_5\, b \ket{ \Lambda_b(p,s_{\Lambda_{b}}) }
        & = -\overline{u}_\Lambda(k,s_{\Lambda}) \:\gamma_5 \bigg[ f_t^A(q^2)\: (m_{\Lambda_b}+m_\Lambda)\frac{q^\mu}{q^2} \\
    \nonumber
        &   \phantom{\overline{u}_\Lambda \bigg[}+ f_0^A(q^2)\frac{m_{\Lambda_b}-m_\Lambda}{s_-}
            \left( p^\mu + k^{ \mu} - (m_{\Lambda_b}^2-m_\Lambda^2)\frac{q^\mu}{q^2}  \right) \\
    \nonumber 
        & \phantom{\overline{u}_\Lambda \bigg[}+ f_\perp^A(q^2) \left(\gamma^\mu + \frac{2m_\Lambda}{s_-} p^\mu - \frac{2 m_{\Lambda_b}}{s_-} k^{ \mu} \right) \bigg]  u_{\Lambda_b}(p_{\Lambda_{b}},s_{\Lambda_{b}}), \\
    \bra{ \Lambda(k,s_{\Lambda}) } \overline{s} \,i\sigma^{\mu\nu} q_\nu \, b \ket{ \Lambda_b(p,s_{\Lambda_{b}}) } 
        &= - \overline{u}_\Lambda(k,s_{\Lambda}) \bigg[  f_0^T(q^2) \frac{q^2}{s_+} \left( p^\mu + k^{\mu} - (m_{\Lambda_b}^2-m_{\Lambda}^2)\frac{q^\mu}{q^2} \right) \\
    \nonumber 
        & \phantom{\overline{u}_\Lambda \bigg[} + f_\perp^T(q^2)\, (m_{\Lambda_b}+m_\Lambda) \left( \gamma^\mu -  \frac{2  m_\Lambda}{s_+} \, p^\mu - \frac{2m_{\Lambda_b}}{s_+} \, k^{ \mu}   \right) \bigg] u_{\Lambda_b}(p,s_{\Lambda_{b}}) \, , \\
    \bra{ \Lambda(k,s_\Lambda) } \overline{s} \, i\sigma^{\mu\nu}q_\nu \gamma_5  \, b \ket{ \Lambda_b(p,s_{\Lambda_{b}}) }
        & = -\overline{u}_{\Lambda}(k,s_{\Lambda}) \, \gamma_5 \bigg[   
        f_0^{T5}(q^2) \, \frac{q^2}{s_-}
            \left( p^\mu + k^{\mu} -  (m_{\Lambda_b}^2-m_{\Lambda}^2) \frac{q^\mu}{q^2} \right) \\
    \nonumber
        &   \phantom{\overline{u}_\Lambda \bigg[}  + f_\perp^{T5}(q^2)\,  (m_{\Lambda_b}-m_\Lambda)
            \left( \gamma^\mu +  \frac{2 m_\Lambda}{s_-} \, p^\mu - \frac{2 m_{\Lambda_b}}{s_-} \, k^{ \mu}  \right) \bigg]  u_{\Lambda_b}(p,s_{\Lambda_{b}})\,,
\end{align}
where we abbreviate $\sigma^{\mu \nu} = \frac{i}{2} [\gamma^\mu, \gamma^\nu]$ and $s_{\pm} = (m_{\Lambda_b} \pm m_\Lambda) - q^2$.
The labelling of the ten form factors follows the conventions of Ref.~\cite{Boer:2014kda}.

\section{Individual background contributions}
\label{app:backgrounds}

Below we list the dominant background sources found in the \BdKstNuNu and \BsPhiNuNu analysis, along with the relative size of each contribution within in each category, after the preselection requirements are implemented.
All of these may come with additional neutrals (\piz, $\gamma$, $\neu$) in the final state.

\subsection{\BdKstNuNu backgrounds with a real $\Kstar(892)^0$}
\begin{table}[H]
    \renewcommand{\arraystretch}{1.2}
    \begin{tabular}{ l c}
        Decay & Relative Size \\
        \hline
        $\Bp \to \Dstarz \neu \ellp$ , $\Dstarz \to \Dz\piz / \gamma$, $\Dz \to \Kstarz \piz / \eta$  & 0.39 \\
        $\Bz \to \Dstarm \neu \ellp$ , $\Dstarm \to \Dz\pim$, $\Dz \to \Kstarz \piz / \eta$ & 0.19  \\
        $\Bp \to \Dz \neu \ellp $, $\Dz \to \Kstarz \piz / \eta$ & 0.11 \\
        $\Bd \to \Dm \neu \ellp$, $\Dm\to\Kstarz \ellm \neub$ & 0.05 \\
        Prompt charm & 0.25
    \end{tabular}
\end{table}

\subsection{\BdKstNuNu backgrounds with a fake $\Kstar(892)^0$}
\begin{table}[H]
    \renewcommand{\arraystretch}{1.2}
    \begin{tabular}{ l c}
        Decay & Relative Size \\
        \hline
        $\Bp \to \Dstarz \neu \ellp$, $\Dstarz \to \Dz\piz/\gamma$, $\Dz\to\hp\hm\piz(\piz)$ & 0.46 \\
        $\Bz \to \Dstarm \neu \ellp$, $\Dstarm \to \Dz\pim$, $\Dz\to\hp\hm\piz(\piz)$ & 0.29 \\ 
        $\Bp \to \Dz \neu \ellp$, $\Dz\to\hp\hm\piz(\piz)$ & 0.17 \\ 
        $\Bd \to \Dm \neu \ellp$, $\Dz\to\hp\hm\piz(\piz)$ & 0.01 \\
        Prompt charm & 0.07
    \end{tabular}
\end{table}

\subsection{\BsPhiNuNu backgrounds with a real $\phi$}
\begin{table}[H]
    \renewcommand{\arraystretch}{1.2}
    \begin{tabular}{ l c}
        Decay & Relative Size \\
        \hline
        $\Bp \to \Dstarz \neu \ellp$, $\Dstarz \to \Dz\piz/\gamma$, $\Dz\to\phi X$ & 0.41 \\
        $\Bz \to \Dstarm \neu \ellp$, $\Dstarm \to \Dz\pim$, $\Dz\to\phi X$ & 0.18 \\ 
        $\Bp \to \Dz \neu \ellp$, $\Dz\to\phi X$ & 0.16 \\ 
        Prompt charm & 0.25
    \end{tabular}
\end{table}

\subsection{\BsPhiNuNu backgrounds with a fake $\phi$}
\begin{table}[H]
    \renewcommand{\arraystretch}{1.2}
    \begin{tabular}{ l c}
        Decay & Relative Size \\
        \hline
        $\Bp \to \Dstarz \neu \ellp$, $\Dstarz \to \Dz\piz/\gamma$, $\Dz\to\hp\hm\piz(\piz)$ & 0.50 \\
        $\Bz \to \Dstarm \neu \ellp$, $\Dstarm \to \Dz\pim$, $\Dz\to\hp\hm\piz(\piz)$ & 0.24 \\ 
        $\Bp \to \Dz \neu \ellp$, $\Dz\to\hp\hm\piz(\piz)$ & 0.20 \\ 
        Prompt charm & 0.06
    \end{tabular}
\end{table}

\section{Comparison between generated and selected signal candidates}
\label{app:kinematic_distributions}

\begin{figure}
    \centering
    \includegraphics[width=0.49\textwidth]{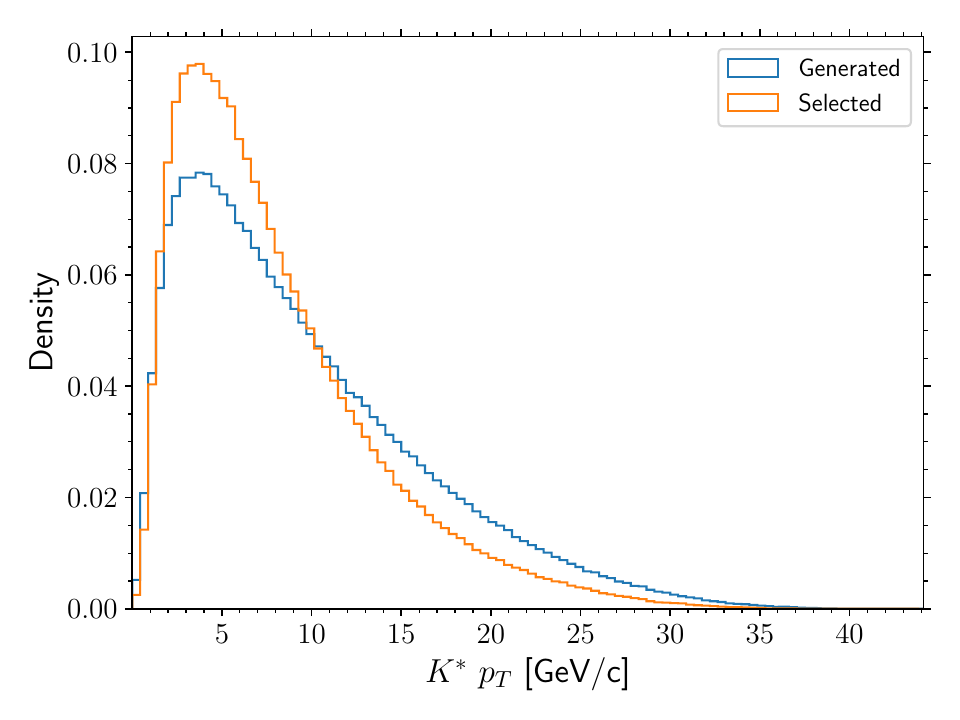}
    \includegraphics[width=0.49\textwidth]{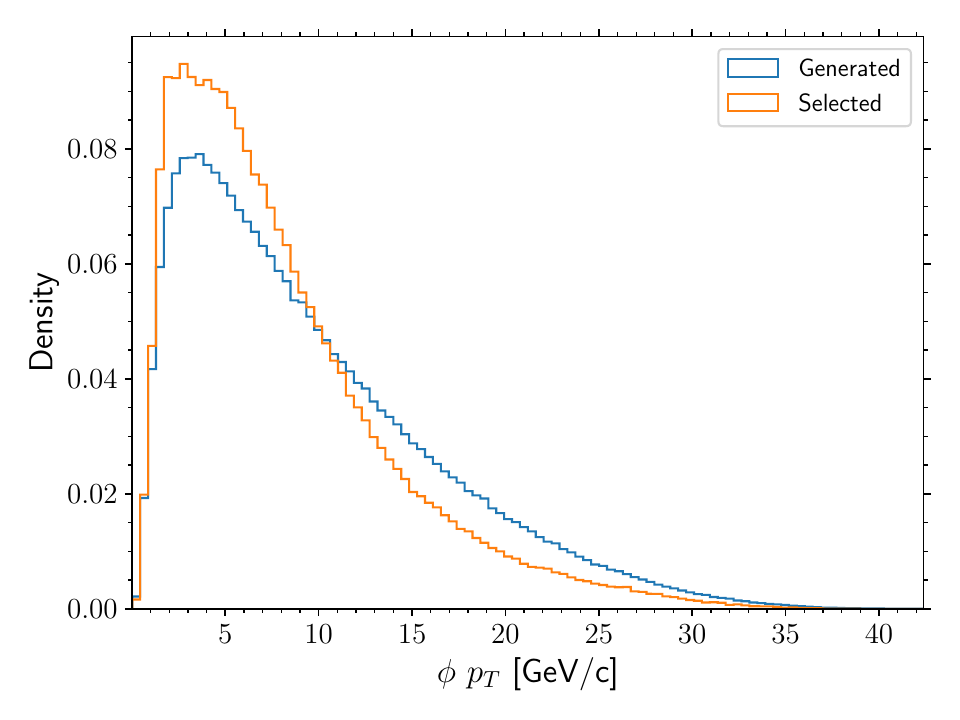}
    \caption{A comparison between the generated (blue) and selected (orange) transverse momentum of \Kstarz (left) and $\phi$ (right) signal candidates.}
    \label{fig:cand_comp_pt}
\end{figure}

\begin{figure}
    \centering
    \includegraphics[width=0.49\textwidth]{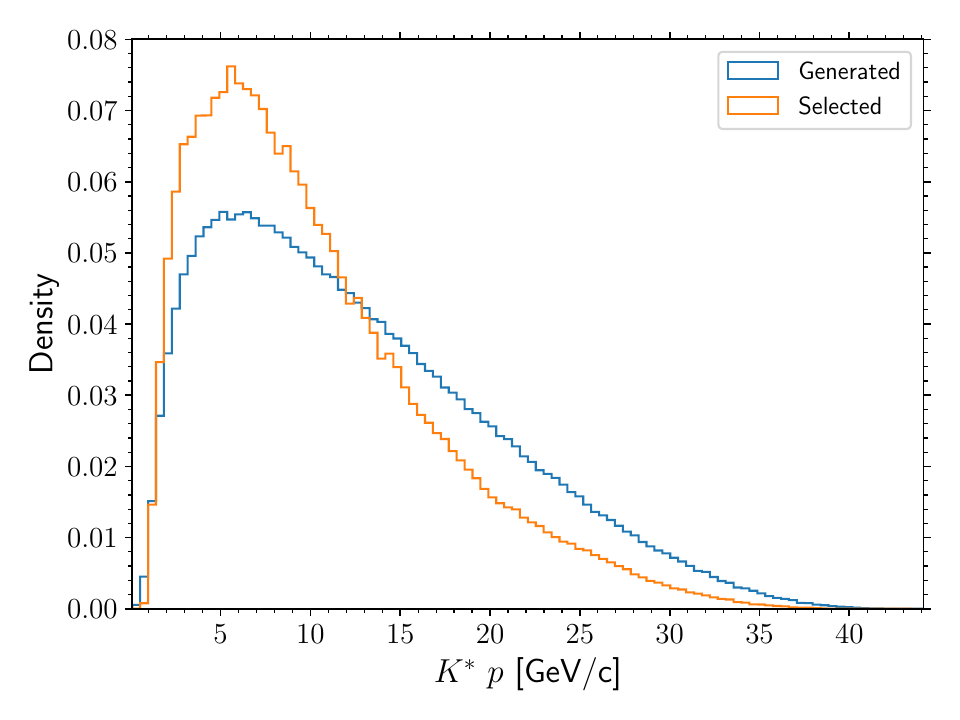}
    \includegraphics[width=0.49\textwidth]{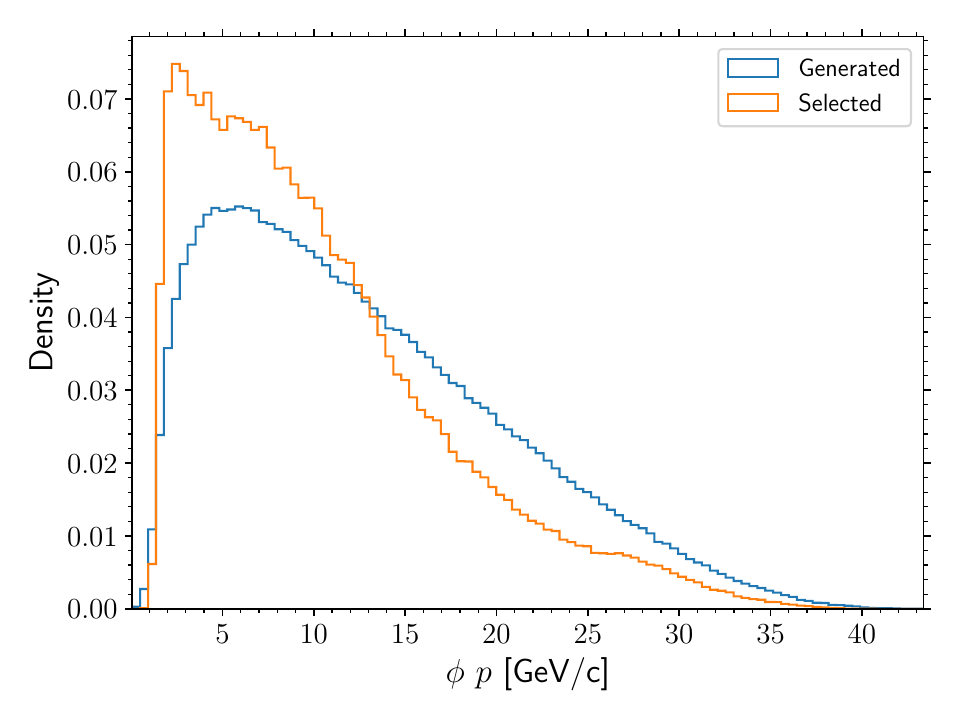}
    \caption{A comparison between the generated (blue) and selected (orange) scalar momentum of \Kstarz (left) and $\phi$ (right) signal candidates.}
    \label{fig:cand_comp_p}
\end{figure}

\begin{figure}
    \centering
    \includegraphics[width=0.49\textwidth]{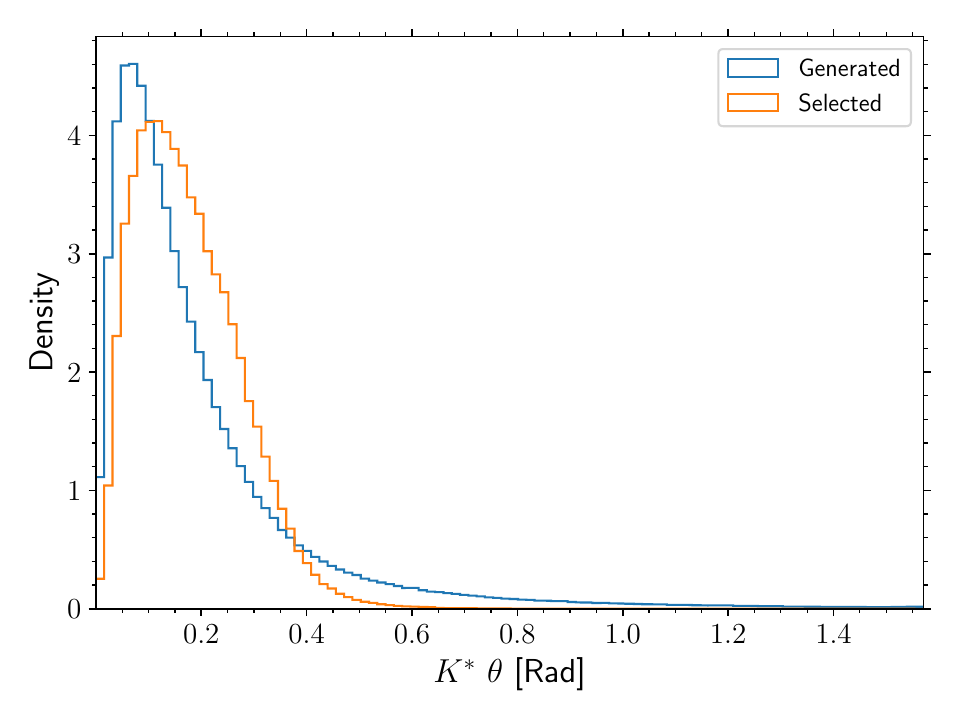}
    \includegraphics[width=0.49\textwidth]{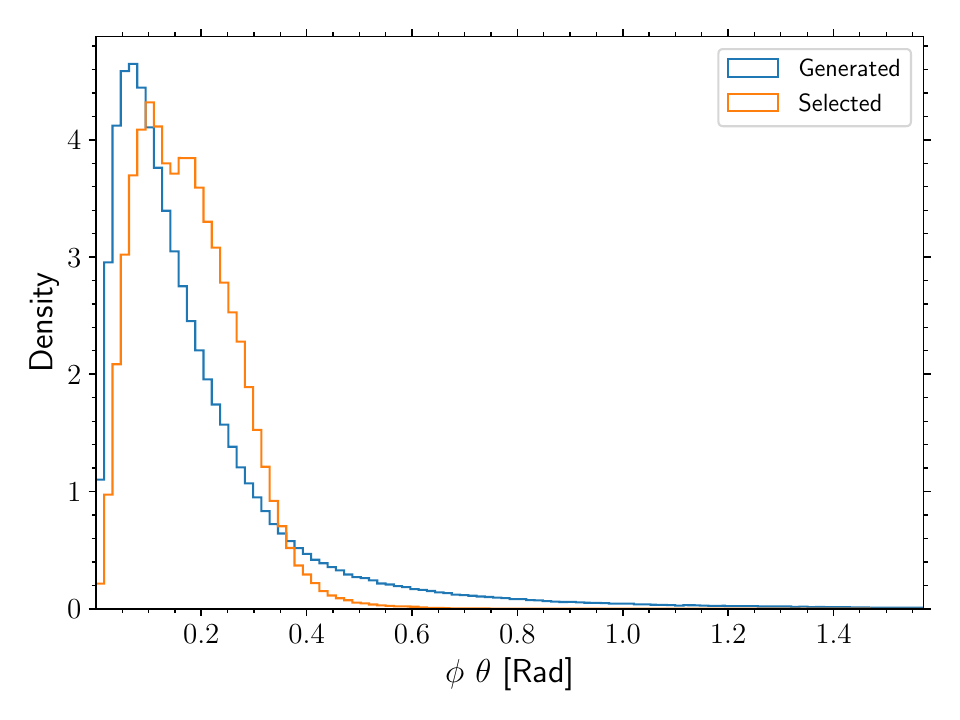}
    \caption{A comparison between the generated (blue) and selected (orange) $\theta$ angle of \Kstarz (left) and $\phi$ (right) signal candidates.}
    \label{fig:cand_comp_theta}
\end{figure}

\begin{figure}
    \centering
    \includegraphics[width=0.49\textwidth]{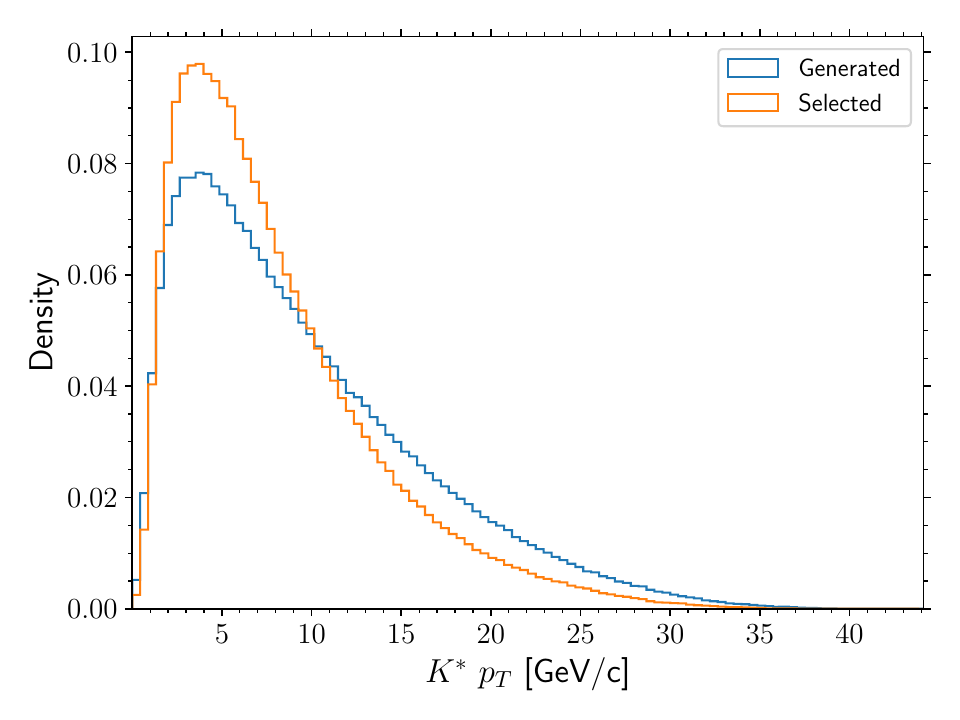}
    \includegraphics[width=0.49\textwidth]{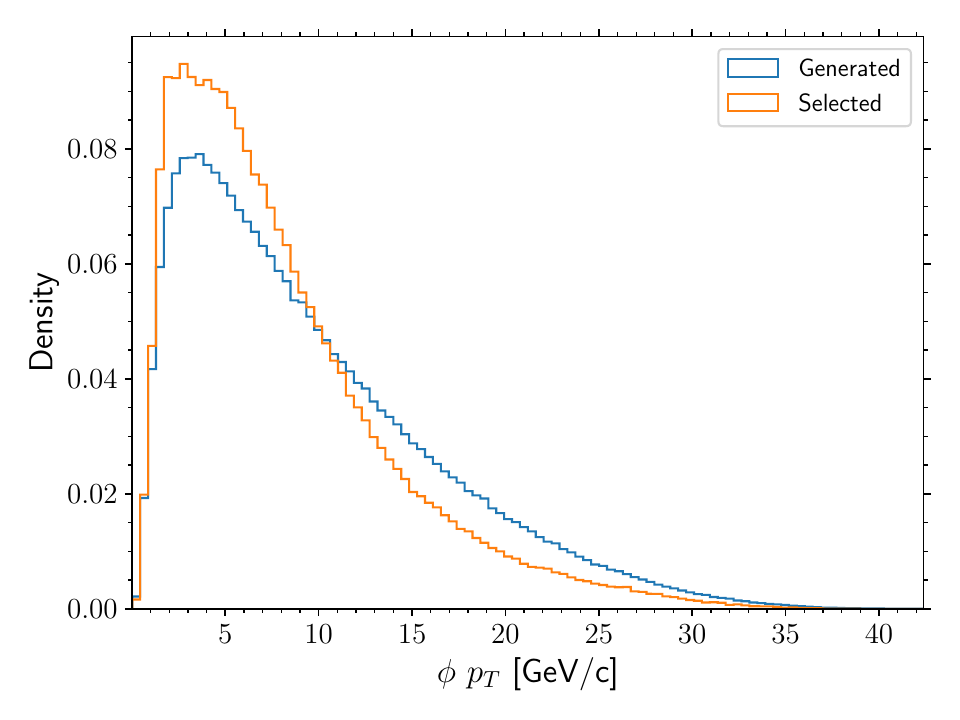}
    \caption{A comparison between the generated (blue) and selected (orange) $\cos(\theta)$ of \Kstarz (left) and $\phi$ (right) signal candidates.}
    \label{fig:cand_comp_ctheta}
\end{figure}

A comparison between generated and selected signal candidates are shown in Fig.~\ref{fig:cand_comp_pt}--~\ref{fig:cand_comp_ctheta}.

\FloatBarrier

\bibliographystyle{LHCb}
\bibliography{references}

\end{document}